\documentclass[twocolumn]{aastex62} 

\usepackage{graphicx}
\usepackage{natbib}

\begin{document}

\title{Stellar Inclination Angles from Be Star H$\alpha$ Emission-Line Profiles}

\correspondingauthor{T.\ A.\ A.\ Sigut}
\email{asigut@uwo.ca}

\author{T.\ A.\ A.\ Sigut}
\affiliation{Department of Physics and Astronomy \\
The University of Western Ontario \\
London, Ontario, Canada N6A~3K7}

\author{A.\ K.\ Mahjour}
\affiliation{Department of Physics and Astronomy \\
The University of Western Ontario \\
London, Ontario, Canada N6A~3K7}

\author{C.\ Tycner}
\affiliation{Department of Physics \\
Central Michigan University \\
Mount Pleasant, MI, USA 48859}

\begin{abstract}

We demonstrate that the angle between star's rotation axis and the observer's line-of-sight, usually called the inclination angle, can be reliably determined for Be stars via H$\alpha$ emission-line profile fitting. We test our method on a sample of 11 Be stars with available long-baseline interferometric data from the Navy Precision Optical Interferometer~(NPOI). We fit the H$\alpha$ emission line profile of each star to obtain a spectroscopic inclination angle $i_{\rm H\alpha}$. We then obtain an independent inclination angle estimate, $i_{\rm V^2}$, by fitting the observed interferometric visibilities with model visibilities based on a purely geometric representation of the light distribution on the sky. The sample differences, $\Delta i \equiv i_{\rm H\alpha} - i_{\rm V^2}$, are normally distributed with a mean of zero and a standard deviation of $6.7$ degrees, and the linear correlation coefficient between $i_{\rm H\alpha}$ and $i_{\rm V^2}$ is $r=0.93$. As Be stars comprise upwards of one fifth of all main-sequence B-type stars, this H$\alpha$ line profile fitting technique has the potential to provide an efficient method for detecting correlated stellar spin axes in young open clusters. Furthermore, if the orientation of the Be star circumstellar disk on the plane of the sky can be constrained by polarization measurements, it is possible to determine the full 3D stellar rotation vector of each Be star.

\end{abstract}

\keywords{stars: rotation - (stars:) circumstellar matter - stars: emission-line, Be - stars: early-type, stars: fundamental parameters}

\section{Introduction}

The angle between a star's rotation axis and the observer's line-of-sight, called the inclination angle $i_\star$, is generally very difficult to observationally constrain. In samples of stars, it is usually assumed that the directional distribution of stellar rotation axes is random, and therefore inclination angles will follow a $\sin i_\star$ distribution for the observer, where $i_\star=0^\circ$ corresponds to the stellar rotation axis pointing directly along the line-of-sight \citep{Gray1992}. Nevertheless, it is important to directly test this common assumption of random orientations as deviations from the $\sin i_\star$ distribution may provide important clues with respect to the interplay among angular momentum, turbulence, and magnetic fields during the formation and evolution of star clusters. Recently, \citet{Corsaro2017} used asteroseismology to measure the inclinations of red giants in two, old, open Galactic clusters~(NGC~6791 and NGC~6819), finding significant stellar spin alignments within these clusters. Furthermore, based on hydrodynamical simulations, \citet{Corsaro2017} suggest that at least half of the initial cluster kinetic energy needs to be in the form of rotation in order for the observed strong spin alignment to be established and be able to persist over the 2--8 Gyr age of these clusters. \citet{Kamann2019} searched for net cluster rotation in both NGC~6791 and NGC~6819, using {\it Gaia\/} data and line-of-sight stellar velocities; they found evidence of systematic rotation for NGC~6791, with an inclination marginally consistent with the alignment found by \citet{Corsaro2017}, but no evidence for cluster rotation in the case of NGC~6819. 

Spin alignment in numerical simulations of star formation has recently been investigated by \citet{ReyRaposo2018}, who examined the role of compressive versus shear turbulence during the early-stages of cluster formation. Like \citet{Corsaro2017}, they concluded that if a significant fraction of the initial kinetic energy is in the form of rotation~($\gtrsim 40$\%), then strong alignment of stellar rotation axes can be produced.

There are several possible approaches to determine the inclination angle $i_\star$ for an individual star. For example, the star's projected rotation speed, $v \sin i_\star$ (where $v$ is the star's equatorial rotational speed) is directly measurable from the rotational broadening of spectral lines \citep{Gray1992}. If repeated observations reveal a periodic variation that can be attributed to the star's rotation, an estimate of the stellar radius can be used to recover $i_\star$ from $v\,\sin i_\star$. The required periodic variations may be provided by magnetic field strength variations or photometric and spectroscopic light variations due to ``star-spots." \cite{Abt2001} used this method on a sample of 102~Ap stars to extract the inclination distribution, finding it random and consistent with the $\sin i_\star$ distribution and showing no correlation with galactic latitude. However, there are disadvantages to this method: a time-series of observations is required to reliably determine the period, an estimate of the stellar radius is required, and other spectral line broadening mechanisms (such as macroturbulence or gravitational darkening -- see below) may complicate the extraction of $v \sin i_\star$ from line profiles. In fact, \cite{Abt2001} found $\sin i_\star>1$ for $\approx\,30$\% of the sample, likely reflecting errors of this type. Recently, \citet{Kovacs2018} has used this method to suggest a non-isotropic distribution of stellar rotation axes in the Praesepe cluster.

Another idea is to recognise that a rotating star cannot be perfectly spherically symmetric due to the additional centrifugal force provided by rotation. A star's equatorial rotational speed is usefully referenced to a ``critical" value, defined as
\begin{equation}
\label{eq:vcrit}
v_{\rm crit} \equiv \sqrt{\frac{GM}{(3/2)\,R_{\rm p}}} \; ,
\end{equation}
at which material at the stellar equator is rotationally-supported \citep{Collins1965}. Here $M$ is the stellar mass, $R_{\rm p}$ is the polar radius, and the factor of $3/2$ in the denominator accounts for the distortion of the stellar surface (in the Roche model) at the critical speed: the equatorial radius is 50\% larger in a critically-rotating star compared to its polar radius. In addition to the geometric distortion, it is well known that a rotating star will have a latitude-dependent effective temperature in which the gas is coolest at the equator and hottest at the poles \citep{vonZeipel1924,Espinosa2011}; near critical rotation, this temperature difference can reach several thousand degrees. In this method, one models spectral lines that are sensitive to this temperature variation in order to extract $v_{\rm frac}\equiv v/v_{\rm crit}$ from spectral synthesis. Combined with the measurement of $v\,\sin i_\star$, one can extract the angle $i_\star$. In practice, self-consistency can be difficult to achieve; for example, the $v\,\sin i_\star$ measurement itself needs to be corrected for gravitational darkening \citep{Townsend2004a}, and the line profile distortion due to gravitational darkening is subtle and high resolution, high signal-to-noise spectroscopic observations are required. In addition, this approach only works for $v_{\rm frac}$ values near unity as the gravitational darkening effects are small for slow and modest rotation. The first application of this method to the analysis of stellar spectra can be found in \citet{Stoe1968}, and more recent application explicitly to the Be stars can be found in \citet{Fremat2005a} and \citet{Zorec2017}.

If a rapidly-rotating and distorted star is both bright and close enough, long-baseline optical interferometry, which directly resolves the stellar surface, can be used to determine the orientation of a star's rotation axis \citep{vanBelle2001}. A good example of this method is interferometric observations of ``spinning-top" star $\alpha\;$Aql or Altair \citep{Domiciano2003a,vanBelle2006,Monnier2007}. For a complete overview, see \citet{vanBelle2012} and references therein.

Another method for determining stellar inclinations is based on asteroseismology. The Fourier transform of the light curves of pulsating stars reveals the frequencies of many non-radial modes of oscillation. \citet{Gizon2003} show that the relative power in rotationally-split, azimuthal modes corresponding to a given angular degree can be used to measure the inclination of the stellar rotation axis. \citet{Corsaro2017}, in their detection of non-random rotation axes orientations in NGC~6791 and 6819, used this method and over four years of {\it Kepler\/} photometric data to accurately determine the pulsation modes of their target red giant stars. A critical evaluation of the asteroseismology method can be found in \citet{Kamiaka2018}, who find that reliable inclinations are possible only from high signal-to-noise time-series data and only in the inclination range $20^\circ \le i_\star \le 80^\circ$.

In this paper, we look at an alternate method applicable to the Be stars and based on the morphology of their H$\alpha$ emission-line profiles. Be stars are rapidly-rotating, B-type main sequence stars that posses an equatorial decretion disk \citep{Rivinius2013}. Although the physics of the disk ejection mechanism remains elusive, disk ejection seems associated with near critical rotation, driven by the internal redistribution of angular momentum within star via rotational mixing \citep{Granada2013a}. The defining observational criteria for Be stars is emission in the hydrogen Balmer series, most notably H$\alpha$ \citep{Slettebak1982}. A wide range of emission line profile morphologies can result from the same disk size and density by varying the viewing angle of the system, from a singly-peaked emission line, to a doubly-peaked line, to a doubly-peaked line with deep shell absorption \citep{Porter2003a}. While the Be stars are a ``peculiarity" class among early-type main sequence stars, they are very common, typically accounting for $\approx 20$\% of all main sequence B-type stars \citep{Zorec1997}. They are also common among Galactic open clusters, with some clusters containing upwards of 40\% of Be stars \citep{Tarasov2012,Tarasov2017}. Extensive surveys of upper main sequence stars, from Galactic to LMC and SMC surveys, include large populations of Be stars \citep{Martayan2006,Martayan2007,Martayan2010,Dunstall2011}.

In the following sections, we show that matching an observed H$\alpha$ line profile to computed profile libraries can be used to reliably estimate $i_\star$ from a single, moderate resolution, moderate signal-to-noise spectrum. We note that this conclusion is contrary to earlier work by \citet{Silaj2010a} who found that the inclination angle {\it could not} be uniquely extracted from the H$\alpha$ profiles of Be stars; however, we trace this difference to the many simplifying approximations made by \citet{Silaj2010a} to compute the H$\alpha$ line profiles. This point is further discussed in the next section.

\section{The H$\alpha$ Line Profile Library}
\label{sec:library}

A large library of H$\alpha$ profiles was computed for Be stars using the \texttt{Bedisk} \citep{Sigut2007} and \texttt{Beray} \citep{Sigut2011a} suite of codes \citep[see also][]{Sigut2018}. For the central, main sequence B-type star, stellar masses between $3.0$ and $20$ M$_{\odot}$ were considered. Radii, luminosities, and effective temperatures for this mass range were determined from the available solar metallicity Geneva evolutionary models of \citet{Ekstrom2012a} corresponding to a central hydrogen fraction $X=0.3$ (approximately the middle-age main sequence). For simplicity, non-rotating models were chosen; this choice is discussed more carefully below as Be stars are known to be rapid rotators \citep{Townsend2004a,Rivinius2013}. The adopted stellar parameters for the central B stars are given in Table~\ref{tab:centralB}.

\texttt{Bedisk} computes the radiative equilibrium temperatures in the Be star's circumstellar disk, given the central star's photoionizing radiation field and the density structure of the disk. The form of the disk density in cylindrical co-ordinates $R$ (distance from the stellar rotation axis) and $Z$ (height above or below the equatorial plane) was taken to be
\begin{equation}
\rho(R,Z) = \rho_0 \, \left(\frac{R_*}{R}\right)^n \, e^{-(Z/H)^2} \,.
\end{equation}
Here $R_*$ is the stellar radius, $\rho_0$ and $n$ are model parameters (see below), and $H$ is the disk scale height. For a disk in vertical, hydrostatic equilibrium at temperature $T_0$, the scale height is given by
\begin{equation}
\frac{H}{R}= \frac{c_s(T_0)}{V_{\rm K}(R)}\,
\end{equation}
where $c_s$ is the sound speed at $T_0$ and $V_{\rm K}(R)$ is the Keplerian orbital speed at $R$. We take $T_0$ to be 60\% of the central star's $T_{\rm eff}$. Note that $T_0$ is used only to fix the scale height; the temperature structure of the disk is found by enforcing radiative equilibrium as noted above \citep[see][for details]{Sigut2009}.

For each B star mass, 15 values of $\rho_0$ between $10^{-12}$ and $10^{-10}\;\rm g\,cm^{-3}$ were considered, along with 11 $n$ values between 1.5 and 4.0 in steps of $0.25$. A \texttt{Bedisk} model was computed for each of the 165 possible density models specified by the various $(\rho_0,n)$ combinations. The \texttt{Bedisk} hydrogen level populations were then used by \texttt{Beray} to compute individual H$\alpha$ line profiles. \texttt{Beray} solves the radiative transfer equation along a large number of rays passing through the Be star+disk system and directed at a distant observer. Rays that terminate on the stellar surface use an appropriately Doppler-shifted, photospheric H$\alpha$ absorption line profile as the boundary condition to the transfer equation; rays that pass completely through the disk assume zero incident radiation. \texttt{Beray} can produce monochromatic images of the system on the sky and spectral energy distributions and detailed spectra for the spatially unresolved system. The H$\alpha$ line profile calculations add two additional parameters, the outer disk radius ($R_d$) and the viewing inclination $(i_\star)$ of the system. Seven disk radii were considered, from 5 to 65$\;R_*$ in steps of 10$\;R_*$, and ten values of the inclination were considered, $0^\circ$ to $90^\circ$ in steps of $10^\circ$. In total, each central B star mass of Table~\ref{tab:centralB} had a library of 11,550 individual H$\alpha$ line profiles, and the total library over all masses considered had 231,000 profiles.

We note that although \citet{Silaj2010a} utilized \texttt{Bedisk} models, their work was performed prior to the development of the \texttt{Beray} code, and their modelling approach employed a number of simplifying approximations in order to compute the H$\alpha$ line profiles using only the \texttt{Bedisk} output. Emergent intensities were computed only for rays passing vertically through the disk\footnote{These vertical rays through the disk were parallel to the star's rotation axis and were naturally computed by the \texttt{Bedisk} code \citep{Sigut2007} that determines the thermal structure of the disk.} and these intensities were assumed to be valid for {\it all\/} viewing inclinations; intensities were then assigned to various disk sectors and Doppler-shifted by each sector's projected velocity. The total disk emission spectrum was then added to a photospheric absorption H$\alpha$ profile appropriate to the star $(T_{\rm eff},\log g$) and its radius. With these assumptions, the disk and stellar spectra are completely separate, and the star can never be viewed through the disk (which is why \citet{Silaj2010a} restricted the inclination angle range considered in their analysis to be $i\le70^\circ$). Finally, \citet{Silaj2010a} assumed all Be star disks had the same size, $30\,R_*$, which restricts the profile shapes in an artificial way. The \texttt{Beray} code, described above and used in the present work, removes all of these assumptions by performing the full radiative transfer formal solution for the intensity received by a distant, external observer from the star+disk system.

Returning to the present work using \texttt{Beray}, to match an individual observed H$\alpha$ profile, the star's spectral type is used to estimate its mass which selects the particular profile library to use. This process can be improved in some cases when additional information about the star is known, such as an effective temperature and/or surface gravity, or if the star is a known binary. Given the library, a figure-of-merit between each computed profile and the observed one is made. The figure-of-merit, ${\cal F}$, is defined as 
\begin{equation}
\label{eq:F}
{\cal F} \equiv \frac{1}{W}\,\sum_{i=1}^N w_i\,\left|\frac{F_i^{\rm mod} - F_i^{\rm obs}}{F_i^{\rm obs}}\right| \,\times 100\% \,,
\end{equation}
where $F_i^{\rm mod}$ is the flux of the model library line profile (interpolated onto the observed wavelength scale), $F_i^{\rm obs}$ is the observed flux profile, and the sum is over the $N$ wavelengths across the observed line profile. All fluxes were continuum normalized before computing ${\cal F}$. Two weightings were considered, $w_i=1$ and $w_i=|F_i^{\rm obs}-1|$. This latter choice, called as ``core-weighting," weights the line emission peaks more than points close to the continuum. Generally the computed profiles fit observations well; however, strong emission lines often have wings that are wider than the models can produce, perhaps due to the neglect of incoherent electron scattering \citep{Poeckert1979}. In these cases, the ``core-weighting" is appropriate -- see \citet{Sigut2015} for a discussion of this in the case of the star o~Aqr. Finally, $W\equiv \sum_i w_i$.

The best library fit is chosen as the profile which minimizes ${\cal F}$. The corresponding parameters $(\rho_0,n,R_d,i_\star)$ are then used as the starting point for a refinement to further minimize ${\cal F}$ by linearly interpolating profiles between the library grid points. It is these refined parameters that are identified as the ``best-fits" to a given profile. However, it is often the case that a number of model profiles will fit any given observed emission line almost equally well; therefore, an uncertainly in each fitted parameter is found by selecting all library profiles with figure-of-merits that satisfy ${\cal F_{\rm rel}}\le 1.15$ where
\begin{equation}
{\cal F}_{\rm rel} \equiv \frac{\cal F}{{\cal F}_{\rm min}} \;.
\label{eq:Frel}
\end{equation}
Here ${\cal F}_{\rm min}$ is the minimum figure-of-merit found by the refinement procedure. While the value of $1.15$ is arbitrary, the profiles selected are quite close upon visual inspection. Thus each fitted parameter, $(\rho_0,n,R_d,i_\star)$, has a best estimate (the refined value) and an uncertainty taken to be the standard deviation of that parameter over library profiles satisfying the ${\cal F}_{\rm rel}\le 1.15$ criteria.

Given the H$\alpha$ line profile library, the first issue is to demonstrate that reliable estimates of inclination can be extracted from a single, observed H$\alpha$ profile despite the fact that (1) the four parameters $(\rho_0,n,R_d,i_\star)$ must be simultaneously determined, (2) the assumed central B star model may be inaccurate, and (3) observations have profiles with finite SNR and spectral resolution constrained by the resolving power ${\cal R}$. This is done in the next section, Section~\ref{sec:sim}, using simulated ``observed" H$\alpha$ line profiles. After this, in Section~\ref{sec:npoi}, an interferometric sample of Be stars with independently determined inclinations is used to put the H$\alpha$ inclinations to the observational test. 

\begin{table}
\begin{center}
\caption{Adopted stellar parameters for the central B stars.\label{tab:centralB}}
\begin{tabular}{llll}\\ \hline\hline
$T_{\rm eff}$ & Mass & Radius & Luminosity\\
(K) & ($M_{\odot}$)  & ($R_{\odot}$) & ($L_{\odot}$) \\ \hline
 11000 & 3.00 & 2.9 & 1.12e+02 \\
 11600 & 3.25 & 3.1 & 1.53e+02 \\
 12200 & 3.50 & 3.2 & 2.04e+02 \\
 12800 & 3.75 & 3.3 & 2.67e+02 \\
 13400 & 4.00 & 3.5 & 3.44e+02 \\
 14000 & 4.25 & 3.6 & 4.31e+02 \\
 14400 & 4.50 & 3.7 & 5.34e+02 \\
 15000 & 4.75 & 3.8 & 6.54e+02 \\
 15600 & 5.00 & 3.9 & 7.93e+02 \\
 16000 & 5.25 & 4.0 & 9.44e+02 \\
 16400 & 5.50 & 4.1 & 1.12e+03 \\
 17000 & 5.75 & 4.2 & 1.31e+03 \\
 17400 & 6.00 & 4.3 & 1.52e+03 \\
 18200 & 6.50 & 4.5 & 2.03e+03 \\
 19200 & 7.00 & 4.7 & 2.65e+03 \\
 20000 & 7.50 & 4.9 & 3.37e+03 \\
 20600 & 8.00 & 5.1 & 4.23e+03 \\
 21400 & 8.50 & 5.2 & 5.19e+03 \\
 22000 & 9.00 & 5.4 & 6.28e+03 \\
 22800 & 9.50 & 5.6 & 7.50e+03 \\
 23400 & 10.0 & 5.7 & 8.88e+03 \\
 25600 & 12.0 & 6.4 & 1.58e+04 \\
 27400 & 13.9 & 7.0 & 2.51e+04 \\
 29000 & 15.9 & 7.7 & 3.69e+04 \\
 30200 & 17.9 & 8.3 & 5.13e+04 \\
 31400 & 19.8 & 8.8 & 6.79e+04 \\ \hline
\end{tabular}\\
\end{center}
\vspace{0.05in}
Notes.- All entries from \citet{Ekstrom2012a} and correspond to a hydrogen core fraction of $X=0.3$.
\end{table}

\section{Potential Degeneracy in Disk Parameters}
\label{sec:sim}

It is first necessary to demonstrate that it is possible to recover more-or-less unique inclinations from H$\alpha$ line-profile fitting. This is not obvious because each fit requires the simultaneous determination of four parameters: $(\rho_0,n,R_d,i_\star)$\footnote{This assumes the fundamental parameters of the central B-type star are exactly known, which is obviously not the case. This issue is addressed in the simulated samples of Sections~\ref{sec:sample2} and \ref{sec:sample4}.}. Often, a range of models can fit a profile equally well, and it is possible that a wide range of inclinations may be selected. 

\begin{figure}
\centering
\includegraphics[width=0.9\columnwidth]{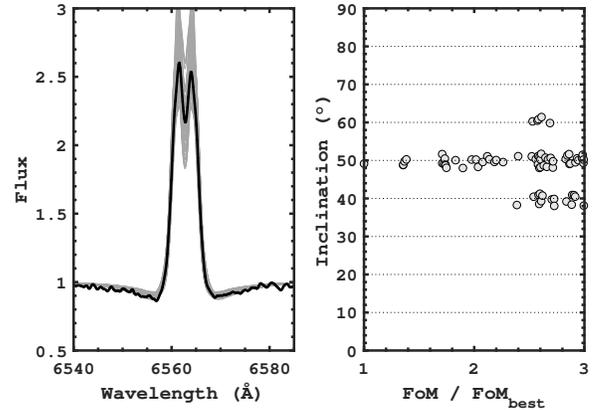}
\caption{Left panel: the dark black line shows a simulated H$\alpha$ line profile for an $M=4.75\,M_{\odot}$ B star surrounded by a disk with parameters $\rho_0=7.69\times 10^{-11}\;\rm g\,cm^{-3}$, $n=3.0$, $R_d=25\,R_*$, and $i_\star=50^\circ$. The simulated profile has ${\cal R}=10^4$ and $\rm SNR=10^2$. The light grey lines are all library profile fits with ${\cal F}_{\rm rel}\le 3$ (see Equation~\ref{eq:Frel}). The right panel shows the inclination angle of the fitting library profiles as a function of ${\cal F}_{\rm rel}$. \label{fig:single_halpha1}}  
\end{figure}

Consider the left panel of Figure~\ref{fig:single_halpha1} which shows a single, simulated H$\alpha$ profile corresponding to an $M=4.75\,M_{\odot}$ B star surrounded by a disk with parameters $\rho_0=7.69\times 10^{-11}\;\rm g\,cm^{-3}$, $n=3.0$ and $R_d=25\,R_*$ seen at an inclination angle of $i_\star=50^\circ$. The profile has been convolved down to a resolution of ${\cal R}=10^4$ and random Gaussian noise has been added to give a SNR of $\approx\,10^2$; these values are typical of Be star observations. This profile was then fit with the appropriate line library, and all profiles that fit with ${\cal F}_{\rm rel}\le3$ are shown. We choose this very large cut-off in ${\cal F}_{\rm rel}$ to make a point about the inclination distribution of the selected models. As can be see from Figure~\ref{fig:single_halpha1}, profiles with very different peak heights, or central depths, are included in the analysis even though they visually do not match the target profile. In the right panel of Figure~\ref{fig:single_halpha1}, the inclination of each fit library profile is shown as a function of the ${\cal F}_{\rm rel}$ of the fit. As can be seen from this figure, all of the best-fitting profiles have an inclination quite close to $i_\star=50^\circ$, and it is only profiles with ${\cal F}_{\rm rel} \ge 2.5$ that have some models closer to $40^\circ$ or $60^\circ$.

\begin{figure}
\centering
\includegraphics[width=0.9\columnwidth]{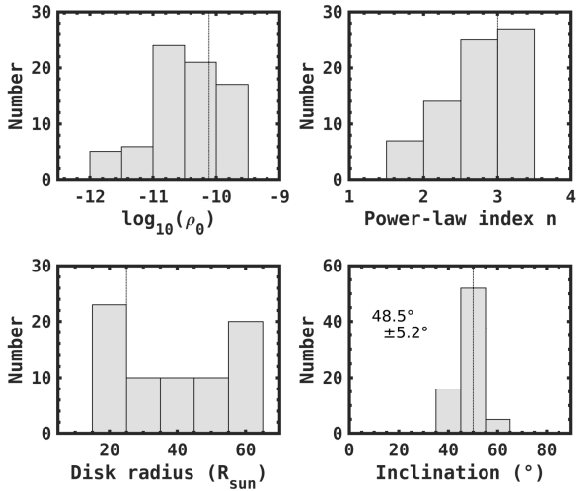}
\caption{Distributions of the recovered model parameters $(\rho_0,n,R_d,i_\star)$ for the fits shown in Figure~\ref{fig:single_halpha1} and satisfying ${\cal F}_{\rm rel}\le 3$. The vertical line in each panel indicates the value of that parameter in the underlying model. \label{fig:single_halpha2}}  
\end{figure}

The distribution of all four disk density parameters $(\rho_0,n,R_d,i_\star)$ for the fits shown in Figure~\ref{fig:single_halpha1} are shown in Figure~\ref{fig:single_halpha2}. While wide distributions are seen in $\rho_0$, $n$ and $R_d$, the distribution in $i_\star$ is a Gaussian of mean $\mu=48.5^\circ$ and standard deviation $\sigma=5.2^\circ$. These distributions are all a result of the Gaussian noise added to create the simulated profile and the (very large) value chosen for ${\cal F}_{\rm rel}$. This result is typical of all of our numerical experiments: while there can be wide distributions in $(\rho_0,n,R_d)$, the inclination angle is well-recovered by the best-fitting profiles. 

To further illustrate the robustness to which the system inclination $i_\star$ can be recovered by H$\alpha$ line-profile fitting, several simulated data-sets of H$\alpha$ line profiles were generated (referred to hereafter as ``samples"). Each sample consisted of 500 members, each representing a star with a randomly assigned mass with a H$\alpha$ profile corresponding to disk density parameters $(n,\rho_0,R_d)$ randomly chosen from within the computed range. The viewing angle $i_\star$ was selected according to the random $\sin i$ distribution. The H$\alpha$ profiles were convolved down to a resolving power of ${\cal R}=10^4$, and random Gaussian noise was added to each profile such that the continuum SNR was $\approx 100$. Finally, each simulated H$\alpha$ line profile was compared to a purely photospheric H$\alpha$ absorption profile (of the same spectral type) using Equation~(\ref{eq:F}) with $w_i=1$. Profiles were rejected if ${\cal F}\le 2$ as these profiles were indistinguishable from a disk-less B star and contained no reliable emission component to constrain $i_\star$. This procedure is necessary as many combinations of $(\rho_0,n,R_d)$ produce essentially no detectable disk emission in H$\alpha$, i.e. those with combinations of small $\rho_0$, large $n$ and small $R_d$. As this selection is principally on $(\rho_0,n,R_d)$, it does not bias the assumed inclination distribution (see below): this is explicitly tested for each sample.

\subsection{Sample 1: No mass errors}
\label{sec:sample1}

The first sample consisted of 500 simulated stars with inclinations and disk parameters chosen as described above. To construct the sample, all masses in Table~\ref{tab:centralB} were assumed equally probable, although in reality the stellar mass function decreases steeply over this range. However, it is also the case that the Be fraction increases modestly with mass, from about 10\% at late spectral types to about 30\% at early spectral types \citep{Zorec1997}.

\begin{figure}
\centering
\includegraphics[width=0.9\columnwidth]{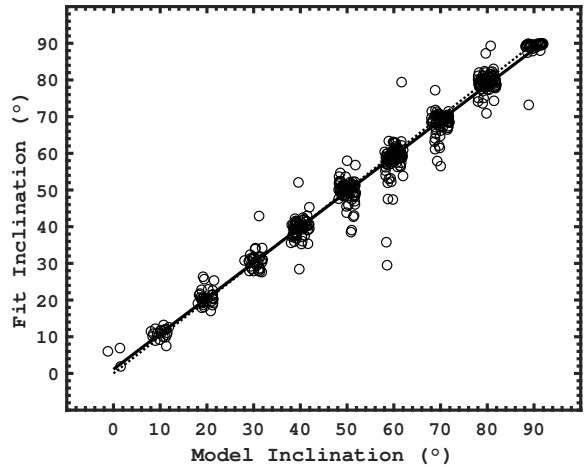}
\caption{Recovered inclinations from H$\alpha$ fitting versus model inclination for Sample~1. The model inclinations range from 0 through 90$^\circ$ in steps of 10$^\circ$; however, the plotted model inclinations are randomly jittered by $\pm 2^\circ$ for clarity. The solid black line is a linear fit to the data, slope $0.97$, and the dotted black line is of unit slope. \label{fig:s1_corr}}  
\end{figure}

\begin{figure}
\centering
\includegraphics[width=\columnwidth]{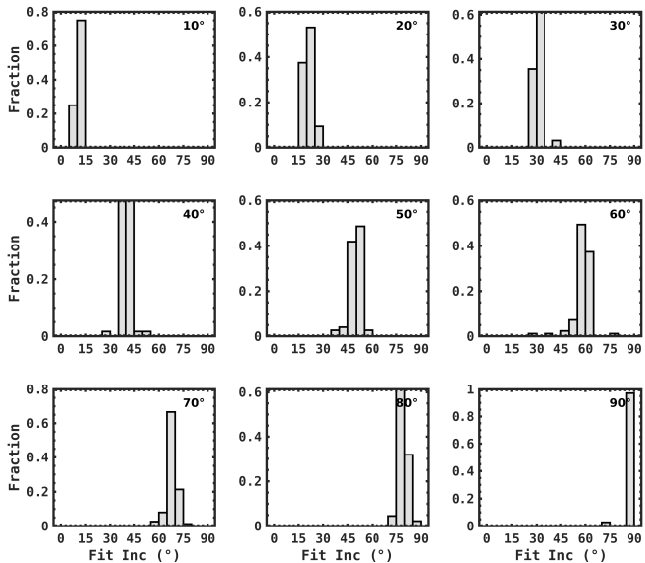}
\caption{Histogram of the inclinations recovered by H$\alpha$ fitting for Sample~1. Each panel is labelled in the top-right by the model inclination used to construct the profile. The model bin $i_\star=0^\circ$, containing only a few stars due to the assumed $\sin i_\star$ distribution (see Figure~\ref{fig:s1_corr}), is not shown for clarity. \label{fig:s1_hist}}  
\end{figure}

\begin{figure}
\centering
\includegraphics[width=0.9\columnwidth]{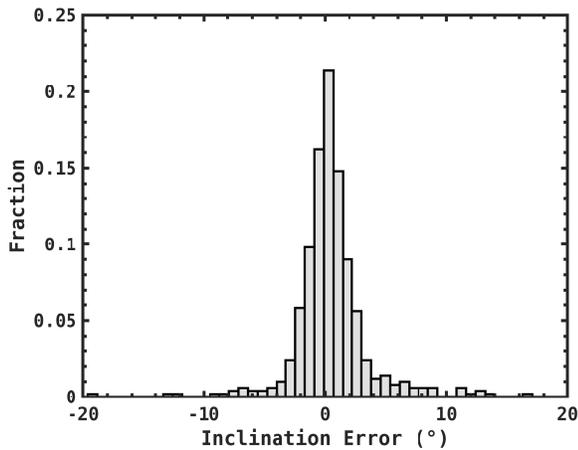}
\caption{Histogram of the inclination error (model minus fit) for all 500 stars of Sample~1. The mean error is $\mu=+0.6^\circ$, and the standard deviation is $\sigma=3.5^\circ$. \label{fig:s1_errhist}}  
\end{figure}

Figure~\ref{fig:s1_corr} plots the recovered inclinations from the H$\alpha$ profile fitting against the model inclination used by \texttt{Beray} to compute the profiles. In this first sample, it is assumed that the parameters of the central B-type star are exactly known; therefore, the profile fitting procedure used to extract the inclination estimate employed the same profile library as used to construct the simulated observed profile. Hence, within the statistical variation of the sample size and the finite SNR and spectral resolution of the simulated profiles, this is an ideal case. As seen in Figure~\ref{fig:s1_corr}, the correlation between the model and recovered inclinations is very strong, with a linear correlation coefficient of $r=0.99$ and a recovered slope of $0.97$. There are a small number of recovered inclinations that differ by larger amounts compared to the model value, and this is more clearly illustrated in Figure~\ref{fig:s1_hist} where histograms of recovered inclinations are shown for each model inclination. Table~\ref{tab:sample1} summarises these results. The standard deviation (or error) in the recovered inclinations peaks at $i_\star=60^\circ$ with $\sigma=6^\circ$; typically the error is $\sigma \le 3^\circ$. The overall error distribution in the recovered inclinations is shown in Figure~\ref{fig:s1_errhist}; the mean of this distribution is $\mu = +0.6^\circ$ and the standard deviation is $\sigma=3.5^\circ$.

A Kolmogorov-Smirnov test (K-S test hereafter)\footnote{As the sample parameter $i_\star$ is binned in discrete steps of $10^\circ$, a simple, binned $\chi^2$ test might be more appropriate; however, the unbinned K-S test is more appropriate to real samples, and none of the results of this section change if the comparison method is switched.} of the recovered inclination distribution versus the random $\sin i_\star$ distribution used to create the sample accepts the null hypothesis that the two distributions are the same. Thus the procedure described in Section~\ref{sec:sim} to exclude from the sample profiles that are too close to the underlying photospheric H$\alpha$ profiles does not bias the inclination distribution.

\begin{table}
\begin{center}
\caption{Recovered inclinations from H$\alpha$ profile fitting for Sample~1.\label{tab:sample1}}
\begin{tabular}{rrrr}\\ \hline\hline
Model~$i_\star$ & \multicolumn{3}{c}{\hrulefill~Fit~$i_\star$~\hrulefill} \\
    & Median & Mean &  $\sigma$ \\ \hline
0   &   6.0 & 4.9  & 2.7 \\
10  &  11.0 & 10.9 &  1.4 \\
20  &  20.4 & 20.7 &  2.2 \\
30  &  30.7 & 30.9 &  2.9 \\
40  &  40.0 & 39.6 &  3.0 \\
50  &  50.1 & 49.7 &  3.1 \\
60  &  59.3 & 58.4 &  5.6 \\
70  &  69.0 & 68.4 &  3.0 \\
80  &  79.5 & 79.5 &  2.4 \\
90  &  89.5 & 88.9 &  2.8 \\ \hline\
\end{tabular}\\
\end{center}
Notes.- All entries are in degrees.
\end{table}

\subsection{Sample 2: Mass errors}
\label{sec:sample2}

The second sample also consisted of 500 simulated stars over the full mass range of Table~\ref{tab:centralB}. The disk parameters and inclinations were chosen as in Sample~1 above; however, the analysis allowed for errors in the stellar masses (or spectral types). The ``analysis" stellar mass ($M_{\rm analysis}$) used to select the H$\alpha$ profile library for the fitting procedure was varied from the original model mass used to compute the profile ($M_{\rm mod}$) according to 
\begin{equation}
\label{eq:mvary}
M_{\rm analysis} = (1+\alpha)\,M_{\rm mod} + \beta\,M_{\rm mod}\,r_{\rm N}(0,1) \;.
\end{equation}
Here $r_{\rm N}(0,1)$ is a Gaussian random deviate of zero mean and unit standard deviation, and $\alpha$ and $\beta$ are parameters that fix a mass offset and mass error, respectively. For Sample 2, $\alpha=-0.2$ and $\beta=0.2$ were chosen; hence, the mass used for the profile analysis was systematically 20\% smaller than the model mass used to construct the profile and had a random variation of 20\% of the model mass (i.e. $\sigma=0.2\, M_{\rm mod}$). This is a very significant mass offset and error, larger even that one might expect from selecting stellar masses based solely on an average main sequence spectral-type-mass calibration. Note, however, that there is some evidence that main sequence B star masses are indeed overestimated by 10 to 20\% \citep{Nieva2014}.  

\begin{figure}
\centering
\includegraphics[width=0.9\columnwidth]{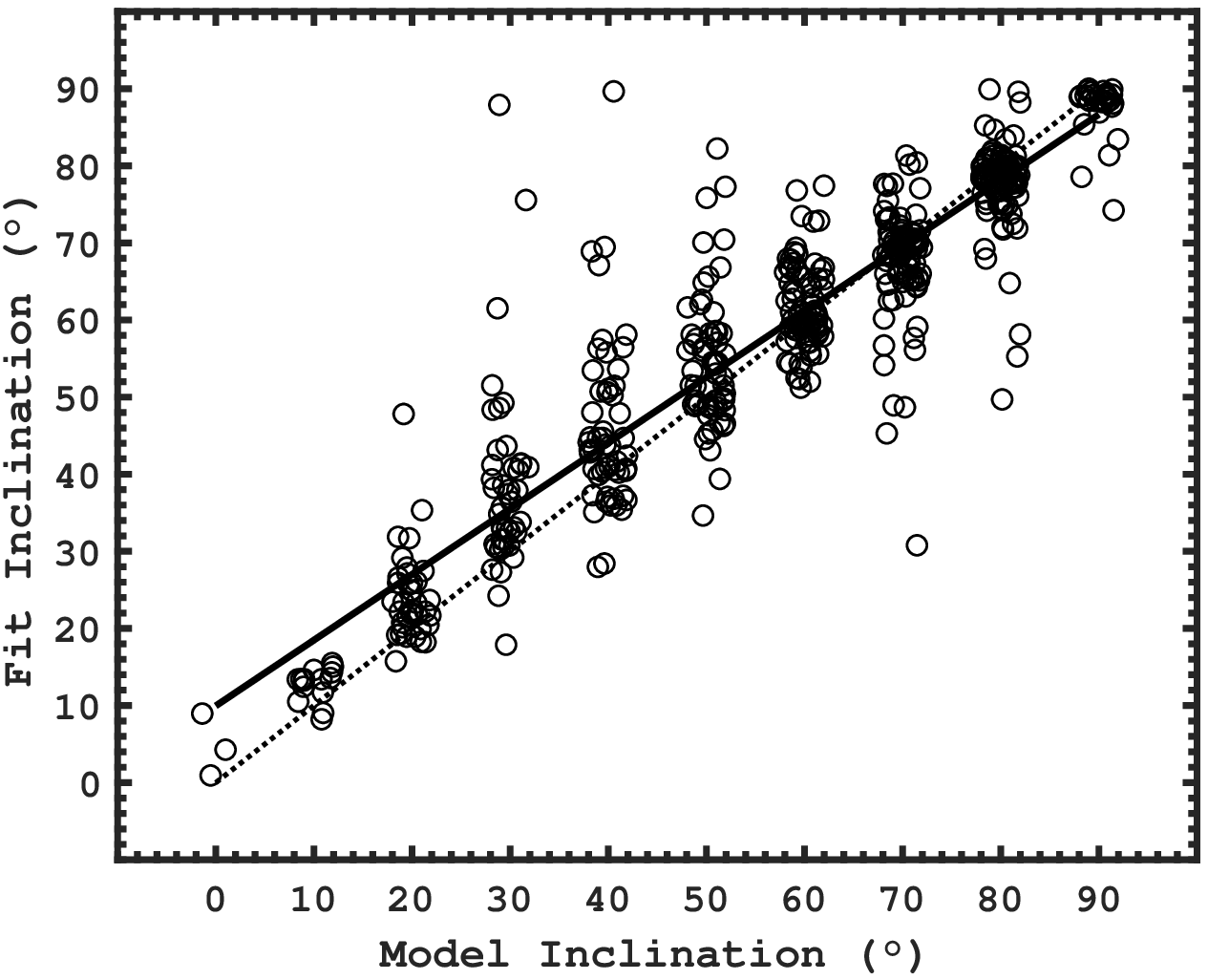}
\caption{Recovered inclinations from H$\alpha$ fitting versus model inclinations for Sample~2. The model inclinations range from 0 through 90$^\circ$ in steps of 10$^\circ$; however, the plotted model inclinations are randomly jittered by $\pm 2^\circ$ for clarity. The solid back line is a linear fit to the data (slope $0.85$) and the dotted black line is of unit slope. \label{fig:s2_corr}}  
\end{figure}

\begin{figure}
\centering
\includegraphics[width=0.9\columnwidth]{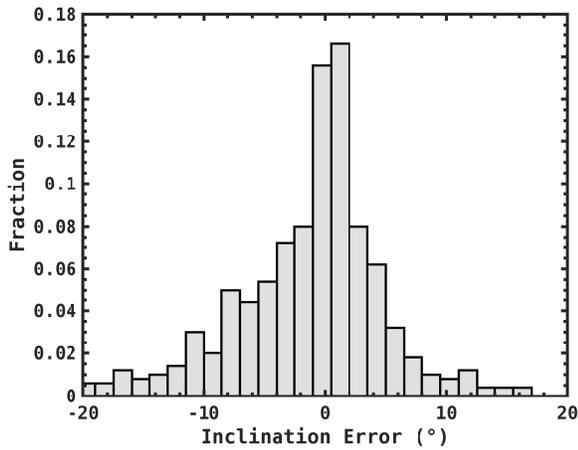}
\caption{Histogram of the inclination errors (model minus fit) for all 500 simulated stars of Sample~2. The mean error is $\mu=-1.6^\circ$, and the standard deviation is $\sigma=8.6^\circ$. \label{fig:s2_errhist}}  
\end{figure}

\begin{figure}
\centering
\includegraphics[width=0.9\columnwidth]{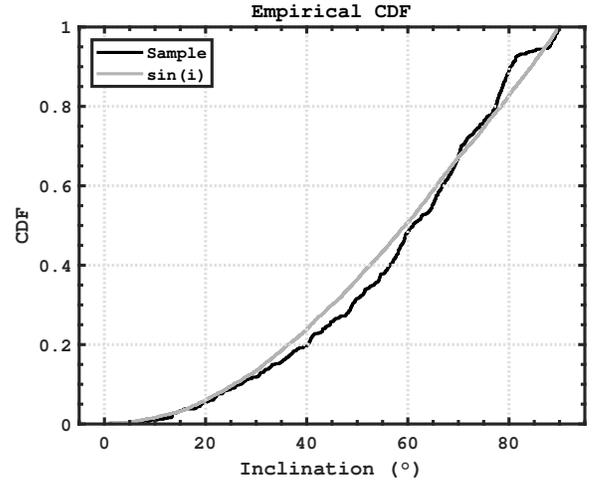}
\caption{Cumulative distribution (CDF) of the recovered fit inclinations for Sample~2 versus the CDF of a random $\sin i$ distribution. A K-S test accepts the null hypothesis that the two underlying distributions are the same. \label{fig:s2_cdf}}  
\end{figure}

Figure~\ref{fig:s2_corr} plots the recovered inclination from the H$\alpha$ profile fits for Sample~2. The error in the recovered inclinations is larger than in the previous sample owing to the large mass errors introduced in the analysis. The correlation between the recovered and model inclinations is still very strong ($r=0.92$), although the recovered slope falls to $0.85$. Figure~\ref{fig:s2_errhist} shows the overall error distribution in the recovered inclinations, and this distribution has a mean error of $\mu=-1.6^\circ$ with a standard deviation of $\sigma=8.6^\circ$. Table~\ref{tab:sample2} sumarizes the uncertainty as a function of model inclination and shows that the error peaks at about $i_\star=30^\circ$ with $\sigma=12.5^\circ$. The cumulative distribution of recovered inclinations is shown in Figure~\ref{fig:s2_cdf}, and a K-S test accepts the null hypothesis that the two distributions are the same.

\begin{table}
\begin{center}
\caption{Recovered inclinations from H$\alpha$ profile fitting for Sample~2.\label{tab:sample2}}
\begin{tabular}{rrrr}\\ \hline\hline
Model~$i_\star$ & \multicolumn{3}{c}{\hrulefill~Fit~$i_\star$~\hrulefill} \\
  & Median & Mean &  $\sigma$ \\ \hline
0  & 4.3  &  4.7  &  4.0 \\
10 & 13.4 & 12.8 & 2.1 \\    
20 & 22.9 & 24.2 & 5.8 \\   
30 & 36.4 & 38.7 & 12.5 \\   
40 & 43.3 & 46.0 & 11.0 \\  
50 & 54.1 & 54.7 & 8.8 \\   
60 & 60.3 & 61.7 & 5.5 \\   
70 & 69.3 & 67.7 & 7.4 \\   
80 & 78.5 & 77.6 & 5.8 \\  
90 & 88.5 & 87.4 & 3.7 \\ \hline
\end{tabular}\\
\end{center}
Notes.- All entries are in degrees.
\end{table}

\section{Observational Test: Comparison to Optical Interferometry}
\label{sec:npoi}

While the results of the previous section are encouraging, only simulated H$\alpha$ profiles were used. In this section, we put the H$\alpha$ fitting method to the observational test. We use the computed H$\alpha$ libraries to fit the observed H$\alpha$ profiles for a sample of 11~Be stars that have available interferometric visibility observations from the Naval Precision Optical Interferometer (NPOI, see \citet{Armstrong1998,Tycner2005}). These observations spatially resolve the circumstellar disk structure on the sky and allow for a completely independent determination of the inclination of the system based on the measured major and minor axes, as discussed below. The sample stars are listed in Table~\ref{tab:NPOI}, along with their spectral types, and the adopted stellar parameters, distances, and references for the visibility data. While none of the interferometric observations are new (and the reader is referred to the individual references in Table~\ref{tab:NPOI}), all of these interferometric visibility observations have been re-analyzed in a uniform way using a bootstrap Monte Carlo method \citep{Wall2003} to estimate the uncertainties in all fit parameters.

Modelling interferometric data allows an estimate of the system viewing inclination that is independent of radiative transfer modeling. As Be star disks are very thin, circular, equatorial disks, the major ($a$) and minor ($b$) axes of the light distribution on the sky must reflect the projection angle $i_\star$. The simplest expectation is that $b=a\cos(i_\star)$. Estimates for $a$ and $b$, and their uncertainties, can be obtained by fitting purely geometric models to the observed interferometric visibilities \citep{Tycner2005}, and this approach is completely independent from the H$\alpha$ line profile libraries above, which are based on radiative transfer models used to derive spectroscopic inclination angles. 

However the relation $i_\star=\cos^{-1}(b/a)$ must fail at some point. While the relative thinness of Be star disks is well established \citep{Porter2003a,Rivinius2013}, they do have a small but finite scale height or opening angle; hence interferometric observations (of sufficient angular resolution) can never yield $b=0$, and it is important to quantify at what ratio $(b/a)$ this simple relation is expected to fail. Appendix~A looks at this issue in detail and concludes that for inclinations $i_\star\le 80^\circ$, $(b/a)$ will yield an accurate inclination. We note that \cite{Cyr2015} looked at this issue in a slightly different way: they examined the statistical distribution of major and minor axes recovered from interferometric observations of Be stars and concluded that the observations best support very thin disks with opening angles of $4^\circ$ to $14^\circ$ in H$\alpha$. 

\begin{table*}
\begin{center}
\caption{Stellar and interferometric characteristics for the 11 Be stars in the NPOI sample\label{tab:NPOI}}
\begin{tabular}{lrlrrr|ccccl}\hline\hline
\multicolumn{6}{c}{\hrulefill\,Stellar Parameters\,\hrulefill} & \multicolumn{5}{c}{\hrulefill\, Interferometry\,\hrulefill} \\
Name       & HD      & Spectral & Mass   & Radius   & Distance & $N_{V^2}$ & $k_{\max}$ & $r=b/a$ & $\phi$ & Ref\\
           &         & Type     & (M$_{\odot}$) & (R$_{\odot}$) & (pc)         &           &(Mcycle/rad)  &  & $(^\circ)$          & \\ \hline
$\gamma$~Cas   &  5394   & B0.5IV & 14.6  & 6.9  & 188      & 169       & 98        & $0.621\pm 0.044$ & $32\pm 2$ & T03 \\
$\phi$~Per     & 10516   & B1.5V  & 11.0  & 5.7  & 221      & 186       & 98        & $0.275\pm 0.010$  & $118\pm 1$ & T06 \\
$\psi$~Per     & 22192   & B5Ve   & 5.5   & 4.65 & 179      & 387       & 94        & $0.323\pm 0.016$ & $133\pm 1$ & S19 \\
$\eta$~Tau     & 23630   & B7III  & 4.2   & 3.2  & 124      & 300       & 57        & $0.839\pm 0.030$ & $40\pm 10$& T05 \\
$48$~Per       & 25940   & B3Ve   & 7.6   & 4.8  & 146      & 291       & 95        & $0.707\pm 0.038$ & $122\pm 5$ & J17 \\
$\beta$~CMi    & 58715   & B8Ve   & 3.8   & 3.0  & 49       & 720       & 120       & $0.695\pm 0.112$ & $140\pm 25$& T05 \\
$\kappa$~Dra   & 109387  & B6IIIe & 4.8   & 6.4  & 140      & 276       & 81        & $0.596\pm 0.065$ & $120\pm 5$ & J08 \\
$\chi$~Oph     & 148184  & B2Vne  & 11.0  & 5.7  & 150      & 132       & 95        & $0.663\pm 0.187$ & $121\pm 31$& T08 \\
$\upsilon$~Cyg & 202904  & B2Vne  & 6.8   & 4.7  & 187      & 201       & 92        & $0.889\pm 0.060$ & $184\pm 51$& J08 \\
$o$~Aqr        & 209409  & B7IVe  & 4.2   & 3.2  & 134      & 994       & 121       & $0.251\pm 0.086$ & $113\pm 4$ & S15 \\
$\beta$~Psc    & 217891  & B6Ve   & 4.7   & 3.6  & 130      & 200       & 92        & $0.810\pm 0.073$ & $133\pm 30$& J08 \\ \hline %
%
\end{tabular}\\
\vspace{0.05in}
Notes.- S19: Sigut et al.\ (2019); J17: \citet{Jones2017}; S15: \citet{Sigut2015}; J08: \citet{Jones2008}; \\ T08: \citet{Tycner2008}; T06: \citet{Tycner2006}; T05: \citet{Tycner2005}; T03: \citet{Tycner2003}
\end{center}
\end{table*}

\begin{figure}
\centering
\includegraphics[width=0.9\columnwidth]{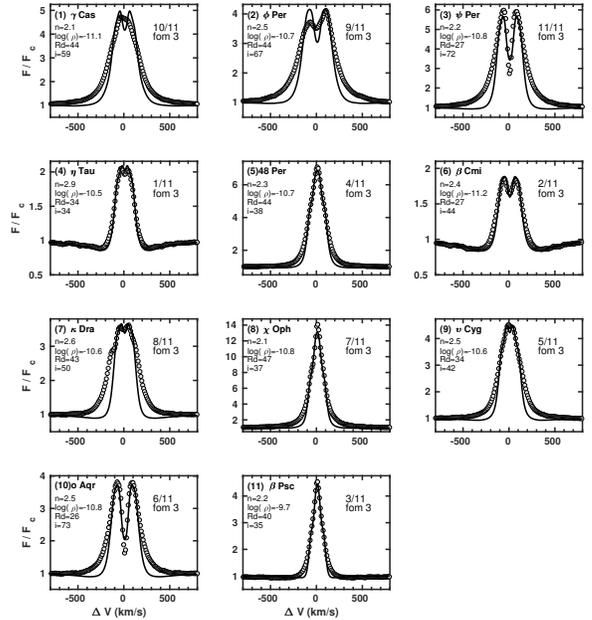}
\caption {Observed (circles) and refined, best-fitting model H$\alpha$ profiles (lines) for each star in the NPOI Be star sample. The name of each star (see Table~\ref{tab:NPOI}) and the best-fit model parameters $(n,\rho_0,R_d,i)$ are given in each panel. The index in the upper right of each panel ranks each fit in the sequence, i.e. $2/11$ is the 2nd best-fit (as measured by $\cal F$ of Equation~\ref{eq:F}) out of the 11 fits. \label{fig:halpha_fit_npi}}  
\end{figure}

Returning to the NPOI Be star sample, Figure~\ref{fig:halpha_fit_npi} shows the H$\alpha$ profiles fits for the 11 stars. All observed profiles were obtained with the John S.\ Hall telescope at Lowell Observatory and have ${\cal R}=10^4$ and SNR of $\approx 10^2$ or better; details for the observations can be found in the references cited in Table~\ref{tab:NPOI}. This sample consists of stars with H$\alpha$ profiles that are nearly symmetric or have small asymmetries in the emission peaks of the profiles ({\it i.e.\/}top section of the H$\alpha$ profile for $\phi\,$Per). Some Be stars can exhibit larger profile asymmetries, and these asymmetries often vary in a cyclic manner on a timescale of several years to a decade \citep{Rivinius2013}. These variations are referred to as $V/R$ variations and are ultimately thought to be the result of a global, one-armed oscillation in the disk's density structure \citep{Okazaki1991}; however, modelling such asymmetries is currently outside the scope of the axisymmetric \texttt{Bedisk} code.

Each observed profile was fit using exactly the same procedure as outlined in Section~\ref{sec:sim} to fit simulated profiles. The wavelength range used to compute the figure-of-merit ${\cal F}$ in Equation~\ref{eq:F} was $\pm 15\,$\AA\ (or $\pm 685\,{\rm km\,s^{-1}}$) of line centre. The adopted spectroscopic inclination, $i_{\rm H\alpha}$, for each star was taken to be the ``refined" inclination, as in Section~\ref{sec:sim}, with an uncertainty taken to be the standard deviation of the inclinations of all library profiles fitting to within ${\cal F}_{\rm rel}\le 1.15$. Overall the resulting fits are fairly good, although there are clear deficiencies in some cases. In particular, there is a tendency for the computed profiles not to be as wide as the observed profiles near the continuum. As noted previously, this could be due to the neglect of incoherent electron scattering in the computed profiles, which can make the profiles wider at their base \citep{Poeckert1979}. For this reason, all fits used core-weighting to compute the figure-of-merit according to Equation~\ref{eq:F} \citep[note that this is the same approach as in][]{Sigut2015}. These 11 profiles fits thus define the H$\alpha$-determined system inclination, $i_{\rm H\alpha}$, for our observational sample.

\begin{figure}
\centering
\includegraphics[width=0.9\columnwidth]{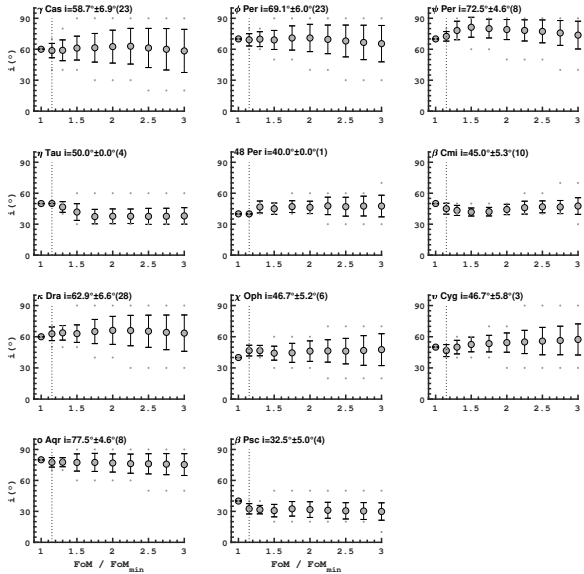}
\caption{The mean and $1\sigma$ error in the spectroscopic H$\alpha$ inclination as a function of ${\cal F}_{\rm rel}$ for the 11 NPOI sample stars. The dots above and below each error bar give the maximum and minimum model inclination at that ${\cal F}_{\rm rel}$. The vertical dotted line in each panel is at ${\cal F}_{\rm rel}=1.15$, and $i_{\rm H\alpha}$ and its error at this value is given next to each star's name; the number is brackets is the number of models satisfying ${\cal F}_{\rm rel}\le 1.15$. \label{fig:coginclination} }
\end{figure}

To gauge the robustness of the inclinations determined by H$\alpha$ profile fitting, we show in Figure~\ref{fig:coginclination} how $i_{\rm H\alpha}$ for the 11 NPOI sample stars varies with the figure-of-merit of the profile fit, in analogy with Figure~\ref{fig:single_halpha1}. The spectroscopic H$\alpha$ inclination is surprisingly flat with ${\cal F}_{\rm rel}$ out to ${\cal F}_{\rm rel}\approx\,3$, although the errors greatly increase. Fits with ${\cal F}_{\rm rel}$ much above 1.5 are visually much poorer fits to the profile, as in Figure~\ref{fig:single_halpha1}.

To model the visibility data for the NPOI sample, we follow \citet{Tycner2005} and fit a purely geometric model to the light distribution on the sky. The model consists of uniform circular disk of fixed angular diameter for the central star and Gaussian elliptical disk for the circumstellar disk contribution, in which case the interferometric signature in the form of squared visibility can be expressed as:
\begin{equation}
\label{eq:v2model}
V^2=\left(\,c_*\,V_*(\theta_*) + (1-c_*)\,V_{\rm GD}(a,b,\phi)\,\right)^2 \;,
\end{equation}
where $\theta_*$ is the fixed angular diameter of the star (based on its assumed radius and distance), and $a$, $b$, and $\phi$ are free model parameters corresponding to the major axis, minor axis, and position angle of the major axis on the sky (N through E), respectively. The parameter $c_*$, also free, is the fractional contribution from the central star to the total flux in the 150~nm wide band-pass centered on H$\alpha$, and is constrained to be $0\le c_* \le 1$.  Detailed forms for the stellar, $V_*$, and disk, $V_{\rm GD}$, visibilities can be found in \citet{Tycner2005}. The uncertainties in the best-fit parameters $(a,b,c_*,\phi)$ were obtained via bootstrap Monte Carlo simulation \citep{Wall2003} using 500 random realizations of the visibility data within the uncertainties. The model fits, with $1\sigma$ uncertainties, for the axial ratio and major axis position angle on the sky are given in Table~\ref{tab:NPOI}.

\begin{figure}
\centering
\includegraphics[width=0.9\columnwidth]{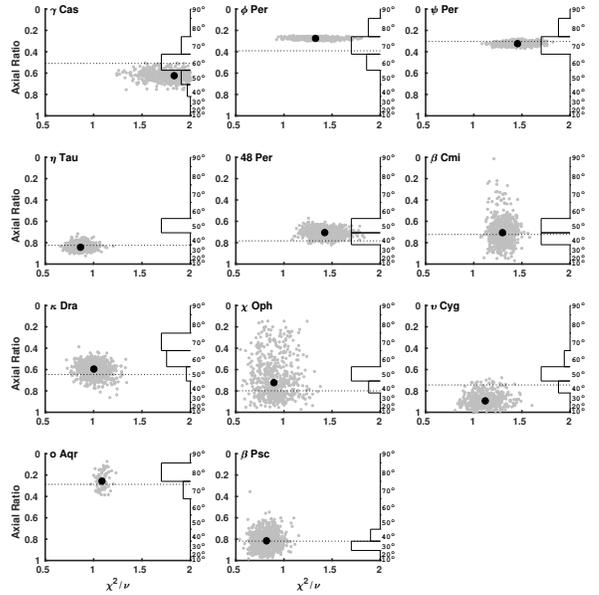}
\caption {Comparison of the interferometric and spectroscopic methods for determining the inclination of the 11 Be stars in the NPOI sample. For each star, the cloud of light grey points is the axial ratio (left vertical axis in each panel) as determined from the visiblities for each of the 500 bootstrap Monte Carlo simulations. These axial ratios are plotted versus the reduced $\chi^2$ of the fit. The median axial ratio is shown as the black circle in each panel. To the left of each panel is a histogram of the inclination found by matching the H$\alpha$ line profile of each star for all models satisfying ${\cal F} \le 1.15 {\cal F}_{\rm min}$, as described in the text. These histograms show the distribution in inclination of the profiles that fit the observations nearly as well as the best-fit profile. Shown as the dotted line in each panel is the inclination of the refined H$\alpha$ line profile fit. Each inclination is assigned a corresponding axial ratio via $r\equiv b/a = \cos i_{\rm H\alpha}$. \label{fig:v2panel}}  
\end{figure}

Comparison of the inclinations obtained from spectroscopic analysis, $i_{\rm H\alpha}$, with those obtained by interferometry, $i_{V^2} \equiv \cos^{-1}(b/a)$, are shown in Figure~\ref{fig:v2panel}. For each star, the interferometrically determined axial ratio is shown as a function of the reduced-$\chi^2$ of the model visibility fit; thus, the interferometric data for each star is represented as a ``cloud" of points for each of the 500 bootstrap Monte Carlo runs. As can be seen from Figure~\ref{fig:v2panel}, the fits all have reduced $\chi^2$ values in the range of 1 to 1.5, indicating that purely geometric model of Equation~\ref{eq:v2model} represents the observed data well. The median axial ratio is shown in the figure, and the cloud of individual points gives a visual depiction of the uncertainty in the axial ratio as determined by interferometry. Also shown in this figure is the axial ratio {\it predicted\/} by the H$\alpha$ line profile fit, namely $\cos(i_{\rm H\alpha})$; the uncertainly in this prediction is represented as a histogram of the inclinations of all model profiles that fit the observed H$\alpha$ emission line within 15\% of the minimum figure-of-merit, i.e.\ ${\cal F}_{\rm rel} \le 1.15$ (see discussion in Section~\ref{sec:library}). As can be seen from Figure~\ref{fig:v2panel}, there is good agreement within the errors of both inclination determination methods, H$\alpha$ line-profile fitting and interferometric modelling. The most discordant case is that of $\eta$~Tau.

\begin{figure}
\centering
\vspace{10pt} 
\includegraphics[width=\columnwidth]{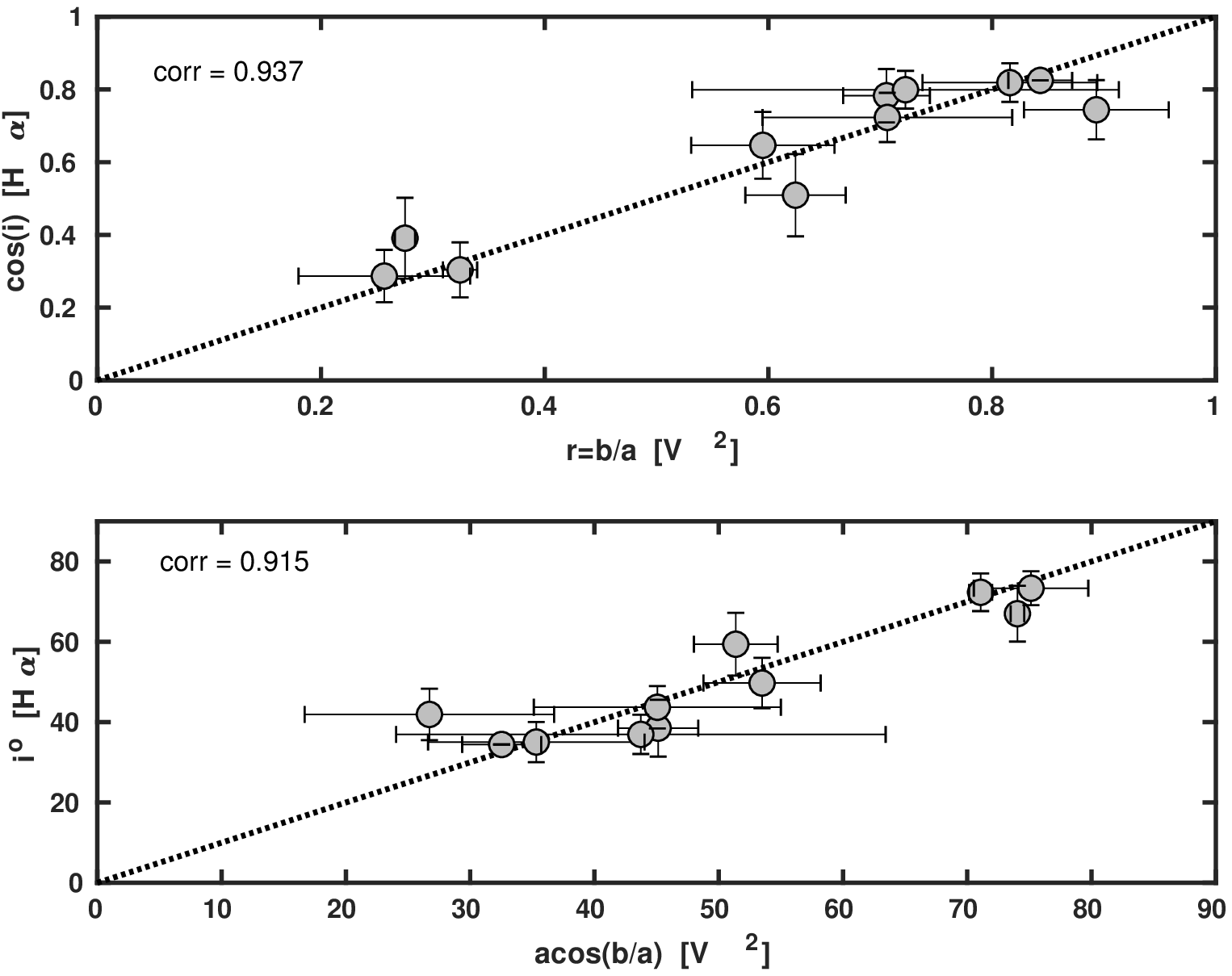}
\caption {Comparison of the axial ratios and system inclinations for both the interferometric and spectroscopic methods. In the top panel, $\cos i_{H\alpha}$ is plotted versus the interferometric axial ratio $r=b/a$. All error bars are 1$\sigma$. In the bottom panel, $i_{H\alpha}$ is plotted versus $\cos^{-1}(b/a)$. The dotted lines in each panel are lines of unit slope and the linear correlation coefficient for the data is noted. \label{fig:v2linear}} 
\end{figure}

Another way to compare the spectroscopic and interferometric results is illustrated in Figure~\ref{fig:v2linear}. In the top panel, the system axial ratio, which is directly measured by interferometry, is compared to value inferred from the spectroscopic H$\alpha$ fit, i.e.\ $\cos(i_{\rm H\alpha})$. In the bottom panel, the system inclination angle, which is directly measured by the spectroscopic H$\alpha$ fit, is compared to the inclination inferred from the interferometric data, namely $i_{\rm V^2}\equiv\cos^{-1}(b/a)$. Error bars ($1\sigma$) are determined via the bootstrap Monte Carlo simulations for the interferometry and the ${\cal F}_{\rm rel} \le 1.15$ inclination histograms for the profile fitting. In both comparisons, the correlation coefficient between $i_{\rm H\alpha}$ and $i_{\rm V^2}$ exceeds 0.9. 

\begin{figure}
\centering
\vspace{10pt} 
\includegraphics[width=\columnwidth]{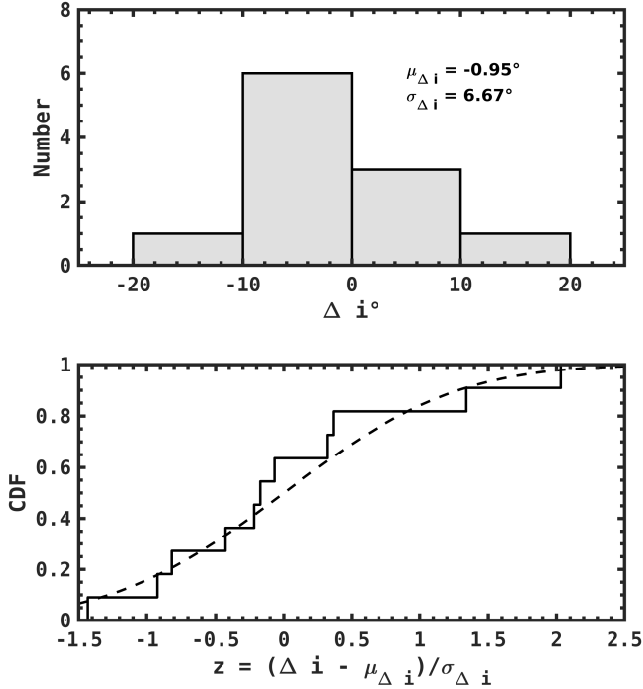}
\caption {Top panel: histogram of $\Delta i_\star \equiv i_{\rm H\alpha} - i_{\rm V^2}$. The sample mean and standard deviation are as indicated. Bottom panel: cumulative distribution of the $z$ statistic corresponding to $\Delta i_\star$ (solid line; defined in the figure) to that of a normal distribution of the same mean and standard deviation (dashed line). A K-S test accepts the null that the two distributions are the same at the 1\% level. \label{fig:ierr}} 
\end{figure}

Figure~\ref{fig:ierr} examines the distribution of inclination errors, defined as $\Delta i_\star \equiv i_{\rm H\alpha} - i_{\rm V^2}$. The upper panel shows a histogram of $\Delta i_\star$; the mean and standard deviation of this distribution are $\mu_{\Delta i}=-0.95^\circ$ and $\sigma_{\Delta i}=6.7^\circ$ respectively. The bottom panel compares the cumulative distribution of the statistic $z \equiv (\Delta i_\star - \mu_{\Delta i})/\sigma_{\Delta i}$ to that of a normal distribution. A K-S tests accepts the null hypothesis that the two distributions are the same at the 1\% level. A direct normal fit to the $\Delta i_\star$ data gives 95\% confidence intervals for the mean of $[-5.4^\circ,3.5^\circ]$ and standard deviation of $[4.7^\circ,11.7^\circ]$; thus, there is no evidence for a systematic difference between the spectroscopic and interferometric determinations of the inclination angle $i_\star$, and the accuracy of the spectroscopic method, relative to the interferometric determination, is about $\pm 7^\circ$, an accuracy well in line with the simulations of Section~\ref{sec:sim}.

\begin{figure}
\centering
\vspace{10pt} 
\includegraphics[width=\columnwidth]{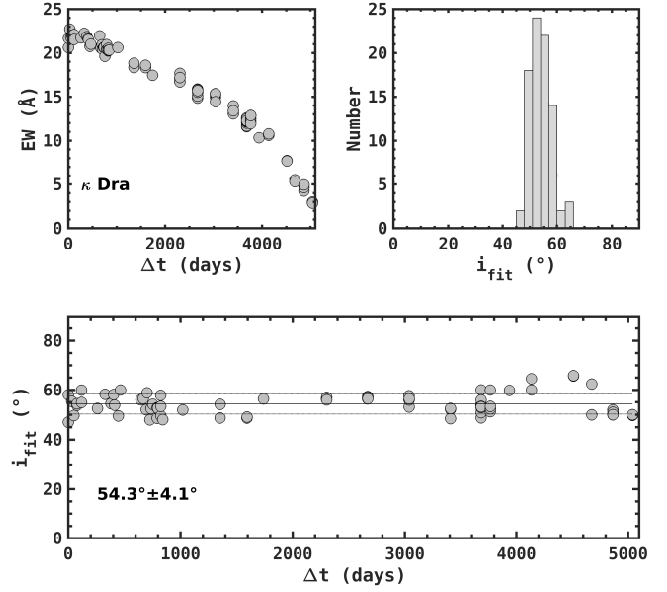}
\caption {The inclination angle of $\kappa\,$Dra obtained from H$\alpha$ spectra obtained over a 15-year period from 2003 to 2018. The upper left panel shows the change in the H$\alpha$ equivalent width (in \AA) over this period. The lower panel shows the system inclination determined from H$\alpha$ line profile fitting as a function of time. The upper right panel shows a histogram of all inclinations, and this distribution is consistent with $i=54^\circ \pm 4^\circ$.  \label{fig:kDra_isum}} 
\end{figure}

Finally, the sample star $\kappa\,$Dra offers another interesting test of using H$\alpha$ profile fitting to determine stellar inclinations. We have spectroscopic observations of the H$\alpha$ line profile of $\kappa\,$Dra covering more than 15 years, from March~2003 through December~2018. During this period, $\kappa\,$Dra's disk has been dissipating, and its H$\alpha$ equivalent width~(EW) has decreased in strength by nearly a factor of five as shown by the EW trend in Figure~\ref{fig:kDra_isum}. We have modelled 85 individual H$\alpha$ line profiles during this time period to extract the viewing inclination as a function of time; of course, the expectation is that the stellar inclination angle is a constant during this period. As shown in Figure~\ref{fig:kDra_isum}, all 15 years of observations are consistent with a constant viewing inclination of $54^\circ\pm 4^\circ$, with no systematic trend in the data. Thus despite the very large changes in the H$\alpha$ line strength and shape, a robust estimate of the viewing inclination can be obtained from any profile.

\section{Recovering \lowercase{non-$\sin i$} Distributions}
\label{sec:nonsini}

Part of the motivation of this work is to develop an efficient technique to search for axial alignments among early-type stars in young, open clusters. In this section, we again turn to samples of simulated H$\alpha$ line profiles constructed with non-$\sin i$ inclination distributions to see if correlations can be recovered under conditions of realistic resolution, signal-to-noise, and sample size.

\begin{figure}
\centering
\includegraphics[width=0.9\columnwidth]{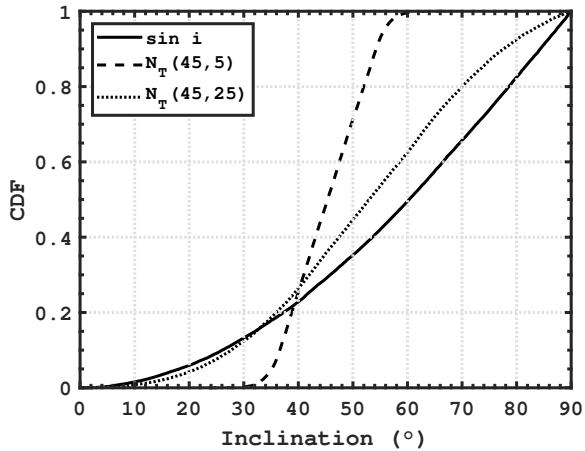}
\caption{Shown are the {\it cumulative\/} distributions of the random $\sin i_\star$ distribution and two truncated Gaussian distributions (Equation~\ref{eq:tg}). The lines are labelled in the legend by their underlying {\it probability\/} distributions. In the legend, $N_T(45,5)$ means Equation~(\ref{eq:tg}) with $i_0=45^\circ$ and $\sigma_0=5^\circ$. \label{fig:egcdf}}  
\end{figure}

For random rotation axis orientations, an external observer will see the familiar $\sin i_\star$ distribution \citep{Gray1992}, i.e.\ the probability of observing an inclination between $i_\star$ and $i_\star+\mathrm{d}i_\star$ is
\begin{equation}
\label{eq:sini}
P(i_\star)\,\mathrm{d}i_\star = \sin i_\star \, \mathrm{d}i_\star \,.
\end{equation}
To simulate samples with preferred axial alignment, we have modified Equation~(\ref{eq:sini}) to be
\begin{equation}
\label{eq:tg}
P(i_\star)\,\mathrm{d}i_\star = {\cal N} \sin i_\star \,N_{\rm T}(i_0,\sigma_0)\, \mathrm{d}i_\star\;.
\end{equation}
Here $N_{\rm T}(i_0,\sigma_0)$ is a truncated Gaussian\footnote{The truncated Gaussian was created with the \texttt{Matlab} \texttt{truncate} function, release R2019a.} restricted to the physical range of $0^\circ\le i_0 \le 90^\circ$ with mean $i_0$ and standard deviation $\sigma_0$. A standard acceptance-rejection method was used to generate inclinations with this distribution \citep{Garcia2000}. This form for $P(i_\star)$ is ad-hoc; however, it is a simple way to parameterize non-random distributions in the simulations to follow. The constant ${\cal N}$ normalizes the distribution and is found via numerical integration given $i_0$ and $\sigma_0$. Examples of these probability distributions are shown in Figure~\ref{fig:egcdf} as {\it cumulative distributions\/}; for example, the cumulative distribution corresponding to $p(i_\star)=\sin i_\star$ is $\mbox{\rm CDF}(i_\star)=1-\cos i_\star$. One desirable feature of Equation~(\ref{eq:tg}) is that it naturally recovers the $\sin i_\star$ distribution in the limit that $\sigma_0$ becomes large.

\subsection{Sample 3: No mass errors}
\label{sec:sample3}

The third sample consists of 500 simulated stars over the full range of masses in Table~\ref{tab:centralB}. No mass errors were considered, but the inclination distribution was taken to be non-random as parameterized by Equation~(\ref{eq:tg}) with $i_0=40^\circ$ and $\sigma_0=20^\circ$. Again H$\alpha$ profiles were generated with SNR~$=10^2$ and ${\cal R}=10^4$, and profiles too similar to pure photospheric H$\alpha$ profiles were excluded.

\begin{figure}
\centering
\includegraphics[width=0.9\columnwidth]{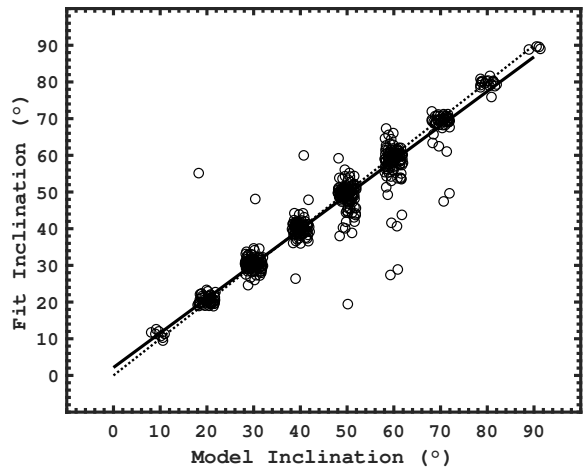}
\caption{Recovered inclinations from H$\alpha$ fitting versus model inclinations for Sample~3. The model inclinations range from 0 through 90$^\circ$ in steps of 10$^\circ$; however, the plotted model inclinations are randomly jittered by $\pm 2^\circ$ for clarity. The solid black line is a linear fit to the data and the dotted black line is of unit slope. \label{fig:s3_corr}}  
\end{figure}

\begin{figure}
\centering
\includegraphics[width=0.9\columnwidth]{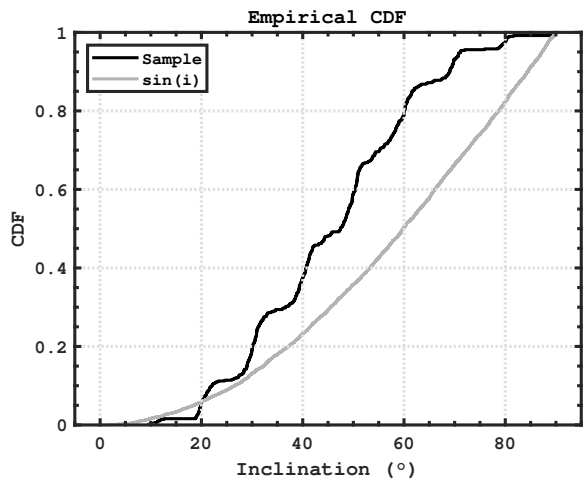}
\caption{Recovered inclinations from H$\alpha$ fitting for Sample~3 as a cumulative distribution. Sample~3 was computed assuming the truncated Gaussian distribution of Equation~(\ref{eq:tg}) with $i_0=40^\circ$ and $\sigma_0=20^\circ$. The random $\sin i$ CDF is also shown. A K-S tests strongly rejects the null hypothesis that the two distributions are the same. \label{fig:s3_cdf}}  
\end{figure}

Figure~\ref{fig:s3_corr} shows the recovered inclinations versus the assumed model inclinations, and Figure~\ref{fig:s3_cdf} shows the recovered inclination distribution expressed as a cumulative distribution. Comparison to the expected CDF of the random $\sin i$ distribution shows very large differences, and a K-S test definitely rejects the null hypothesis that the two distributions are the same. Note that the apparent oscillation in the CDF of the recovered inclinations reflects the discrete values for the model inclinations, which were computed from $0^\circ$ to $90^\circ$ in steps of $\Delta i_\star=10^\circ$.

\begin{figure}
\centering
\includegraphics[width=0.9\columnwidth]{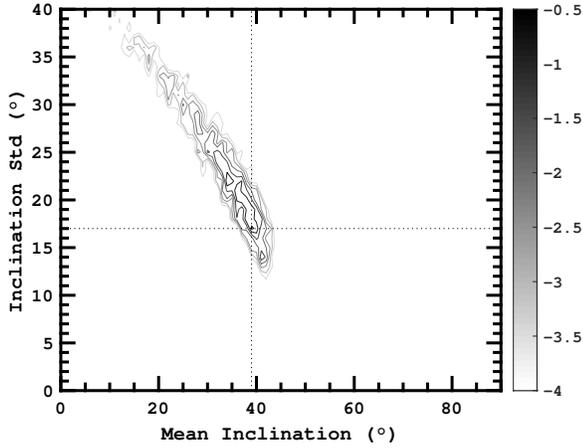}
\caption{Fit in the $(i_0,\sigma_0)$ plane of the recovered H$\alpha$ fit inclinations of Sample~3 to the truncated Gaussian distribution of Equation~(\ref{eq:tg}). The best-fit, as defined by the maximum probability of obtaining a KS-statistic equal to or larger than the observed one, and shown as the intersection of the dotted lines, occurs at $i_0=39^\circ$ and $\sigma_0=17^\circ$ with $\log_{10} (P_{\rm KS})=-0.38$. \label{fig:s3_ksfit}}  
\end{figure}

Figure~\ref{fig:s3_ksfit} is an attempt to recover the parameters used in the truncated Gaussian to construct the sample ($i_0=40^\circ$ and $\sigma_0=20^\circ$) from the recovered inclinations of the 500 simulated stars. For each value in the $(i_0,\sigma_0)$ plane of Figure~\ref{fig:s3_ksfit}, the corresponding truncated Gaussian distribution of Equation~(\ref{eq:tg}) was computed and compared to the recovered inclination sample via a K-S test. Each point in the $(i,\sigma)$ plane was assigned the value $\log_{10} (P_{\rm KS})$, where $P_{\rm KS}$ is the probability of observing a K-S statistic equal to or larger than the one observed. The best-fit model is then the combination of $(i_0,\sigma_0)$ that maximizes this probability. This method formally recovers $i_0=39^\circ$ and $\sigma_0=17^\circ$, maximizing the probability at $\log_{10} P_{\rm KS}=-0.38$; however, there is also a long tail trailing to lower $i_0$ and larger $\sigma_0$. Thus for this sample with no mass errors, there is sufficient information in the recovered inclination distribution to both conclusively rule out the $\sin i_\star$ distribution, and reliably determine the two parameters $(i_0,\sigma_0)$ in the underlying inclination distribution. 

\subsection{Sample 4: Mass errors}
\label{sec:sample4}

In the previous subsection, it was shown that it is possible to recover the parameters in a truncated Gaussian for the inclination distribution. However, this was done for the ideal case of no mass errors. As the results in Section~\ref{sec:sample2} show, the introduction of uncertainty in the underlying central B-type star parameters does reduce the accuracy of the recovered inclinations. This fourth sample consisted of 500 simulated stars over the full range of masses in Table~\ref{tab:centralB} and was analyzed assuming the same large mass errors of Sample~2 ($\alpha=-0.2$ and $\beta=0.2$ in Equation~\ref{eq:mvary}). The inclination distribution was again taken to be Equation~(\ref{eq:tg}) with parameters $i_0=40^\circ$ and $\sigma_0=20^\circ$. Sample H$\alpha$ profiles were generated with SNR~$=10^2$ and ${\cal R}=10^4$, and profiles too similar to pure photospheric H$\alpha$ profiles were excluded.

\begin{figure}
\centering
\includegraphics[width=0.9\columnwidth]{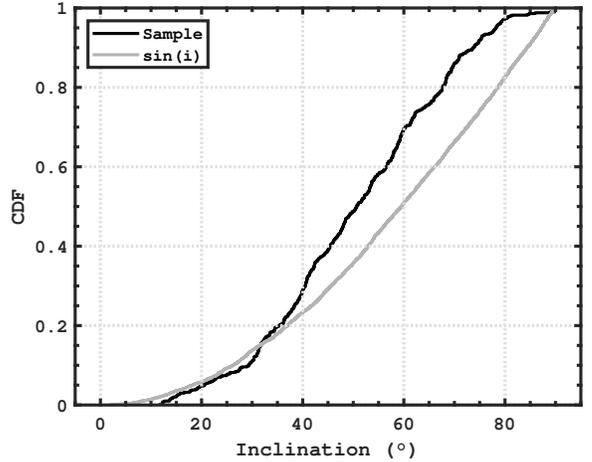}
\caption{Recovered H$\alpha$ fit inclinations for Sample~4 shown as a cumulative distribution. Sample~3 was computed assuming the truncated Gaussian distribution of Equation~(\ref{eq:tg}) with $i_0=40^\circ$ and $\sigma_0=20^\circ$. The random $\sin i$ CDF is also shown. A K-S tests strongly rejects the null hypothesis that the two distributions are the same. \label{fig:s4_cdf}}  
\end{figure}

 Figure~\ref{fig:s4_cdf} shows the recovered inclination distribution expressed as a cumulative distribution. Note that in this case, the oscillations seen in CDF of Sample~4 are absent due to the smoothing influence of the inclination errors introduced by the large mass errors. Comparison to the expected CDF of the random $\sin i$ distribution again shows very large differences, and a K-S test definitely rejects the null hypothesis that the two distributions are the same. 

\begin{figure}
\centering
\includegraphics[width=0.9\columnwidth]{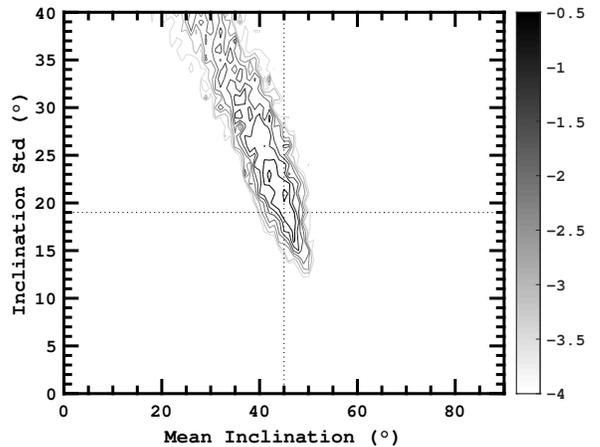}
\caption{Fit in the $(i_0,\sigma_0)$ plane of the recovered H$\alpha$ fit inclinations of Sample~4 to the truncated Gaussian distribution of Equation~(\ref{eq:tg}). The best-fit, as defined by the maximum probability of obtaining a KS-statistic equal to or larger than the observed one, and shown by the intersection of the dotted lines, occurs at $i_0=45^\circ$ and $\sigma_0=19^\circ$ with $\log_{10} (P_{\rm KS})=-0.059$. \label{fig:s4_ksfit}}  
\end{figure}

Figure~\ref{fig:s4_ksfit} attempts to recover the parameters used in the truncated Gaussian to construct the sample ($i_0=40^\circ$ and $\sigma_0=20^\circ$) from the recovered inclinations of the 500 simulated stars. This method formally finds that $i_0=45^\circ$ and $\sigma_0=19^\circ$, maximizing the probability at $\log_{10} (P_{\rm KS})=-0.059$. Again, there is also a long tail trailing to lower $i_0$ and larger $\sigma_0$. Thus, even the large mass errors used in the analysis to extract the inclinations do not prevent a reliable recovery of the parameters in the underlying inclination distribution. 

\subsection{Small Sample Sizes}

The previous sample sizes of 500 Be stars are larger than one might expect in practice; as Be stars comprise approximately one-fifth of all B stars, samples sizes of several thousand early-type spectra would be required. We note that \citet{Corsaro2017}, who reported axial alignment in the old Galactic clusters NGC~6791 and 6819, used samples sizes of $\approx\,$20 red giant stars per cluster. In this section, we re-run our attempted recovery of the parameters in a truncated Gaussian distribution of inclination angles (with and without mass errors) for small sample sizes of $N=15$ and $25$ Be stars. Our results are summarized in Table~\ref{tab:smallN}. 

For each sample size, three different distributions were tested (corresponding to three alignment angles in Equation~\ref{eq:tg}: $i_0=20^\circ$, $40^\circ$ and $70^\circ$), both with and without mass errors, making for a total of 12 random samples.  For the most part, $N=25$ sample sizes reject the null hypothesis of a $\sin i$ distribution and recover the parameters used in the truncated Gaussian, even if samples include large mass errors. However, there were two failures to reject the $\sin i$ distribution corresponding to parameters $(i_0,\sigma_0)=(70^\circ,10^\circ)$ and $(40^\circ,20^\circ)$. In the former case, the peak in the alignment distribution occurred where $\sin i$ was already large, making it hard to distinguish with small $N$. In the latter case, the width of the alignment distribution $(\sigma_0$) was large. Note that these are all single sample realizations, and there is a statistical fluctuation; for example, repeated runs show that the $(70^\circ,10^\circ)$ distribution without mass errors will reject the null $\sin i$ distribution more often than the case with mass errors, contrary to the single set of runs shown in the table. In the second part of Table~\ref{tab:smallN}, we see that $N=15$ samples mostly fail to reject the null hypothesis of a $\sin i$ distribution. In fact, only the $(20^\circ,10^\circ)$ distributions are reliably rejected as the peak occurs where the $\sin i$ distribution is low.

\begin{table}
\begin{center}
\caption{Analysis of non-$\sin i$ distributions using small samples.\label{tab:smallN}}
\begin{tabular}{ccc|cc}\hline\hline
$N_{\rm star}$  & $\Delta\,M^a$  & Distribution  & Null$^b$  & Recovered \\
                & $(\alpha,\beta)$ & $N_{T}(i_0,\sigma_0)$ &    & $N_{T}(i,\sigma)$ \\ \hline
25              & --           & $(20^\circ,10^\circ)$ & R  & $(21^\circ,7^\circ)$ \\  
25              & --            & $(70^\circ,10^\circ)$ & A  &  \\
25              & --           & $(40^\circ,20^\circ)$ & A  &  \\
25              & $(-0.2,0.2)$ & $(20^\circ,10^\circ)$ & R  & $(26^\circ,8^\circ)$ \\
25              & $(-0.2,0.2)$ & $(70^\circ,10^\circ)$ & R  & $(66^\circ,9^\circ)$ \\
25              & $(-0.2,0.2)$ & $(40^\circ,20^\circ)$ & R  & $(43^\circ,18^\circ)$ \\
                &              &                       &    &                      \\
15              & --           & $(20^\circ,10^\circ)$ & R  & $(16^\circ,10^\circ)$ \\
15              & --           & $(70^\circ,10^\circ)$ & A  & \\
15              & --           & $(40^\circ,20^\circ)$ & A  & \\
15              & $(-0.2,0.2)$ & $(20^\circ,10^\circ)$ & R  & $(21^\circ,9^\circ)$ \\
15              & $(-0.2,0.2)$ & $(70^\circ,10^\circ)$ & A  & \\
15              & $(-0.2,0.2)$ & $(40^\circ,20^\circ)$ & A  & \\ \hline
\end{tabular}\\
\vspace{0.05in}
\begin{minipage}[b]{8cm}
Notes.- {\bf a}: Entries are the values used in Equation~\ref{eq:mvary}. ``--" means no mass errors. {\bf b}: The null hypothesis is that the underlying distribution is $\sin i$. An entry of A means the null is accepted, and R, the null is rejected (both at the 5\% significance level).
\end{minipage}
\end{center}
\end{table}

\section{Constraint of Polarization}
\label{sec:polar}

\begin{figure}[htb]
\centering
\vspace{10pt} 
\includegraphics[width=\columnwidth]{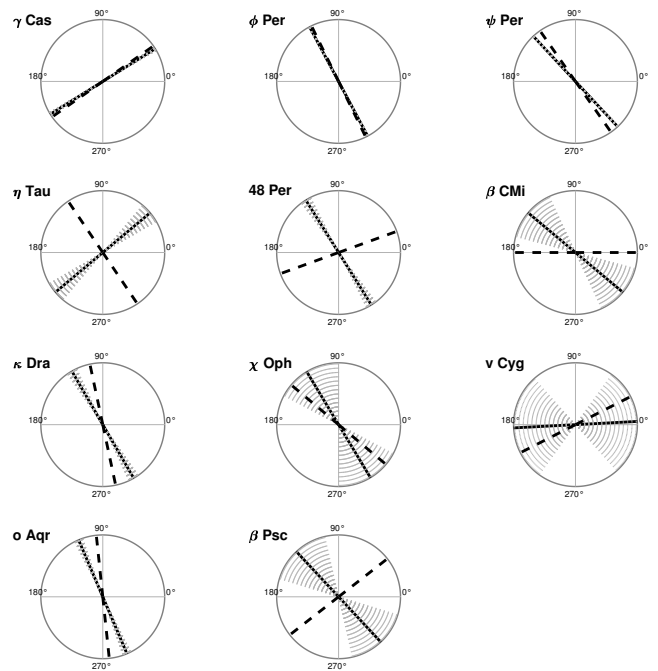}
\caption {Comparison of the position angle of the circumstellar disk on the sky (N through E) as obtained from interferometry (solid line) and polarization (dotted line) for each of the 11 Be stars in the NPOI sample taken from \citet{Yudin2001}. The shaded region in each circle gives range of the uncertainty in the interferometric position angle.\label{fig:phichi}}
\end{figure}

Knowing the system inclination constrains the central B-type star's rotation axis to lie within a cone of opening angle $i_\star$ relative to the line of sight (with the additional $180^\circ$ ambiguity of whether it is the north or south stellar pole facing the observer). The component of the star's rotation axis in the plane of the sky remains unknown. Fortunately, there is an additional observed property of Be stars that potentially constrains the in-sky component, namely continuum polarization. The integrated light of Be stars is known to be weakly polarized due to electron scattering in the flattened circumstellar disk \citep{Yudin2001,Rivinius2013}. The plane of polarization will be perpendicular to the scattering plane, i.e.\ the circumstellar disk, and therefore the polarization position angle on the sky will be perpendicular to the plane of the disk \citep{Brown1977,Poeckert1977}. As the NPOI Be star sample in this work yields the position angle of the disk on the sky ($\phi$, see Table~\ref{tab:NPOI}), we can test this simple prediction, following \citet{Quirrenbach1997}. \cite{Yudin2001} gives a large catalogue of Be stars with intrinsic polarization percentages and position angles ($\chi$), corrected for interstellar polarization (see below).  The angle $\chi+90^\circ$ can be compared to $\phi$ for each star, and this is done in Figure~\ref{fig:phichi} for the 11 NPOI sample Be stars. As there is an $180^\circ$ ambiguity in both of these angles, they are represented in Figure~\ref{fig:phichi} in the unit circle. The interferometric analysis above also provides uncertainties in the interferometric position angles via the Monte Carlo boot-strap analysis, and these are also shown in the figure. Unfortunately, there are no quoted uncertainties for the polarization angles $\chi$.

As can be seen from Figure~\ref{fig:phichi}, five of eleven stars have agreement of $\phi$ and $\chi+90^\circ$ to within the interferometric uncertainties alone, and a further three are quite close and potentially agree for any reasonable errors in the polarization position angle. However, there are three significant disagreements: $\beta$~Psc, $\eta$~Tau and 48~Per, and in these cases, the angle $\chi+90^\circ$ {\it seems\/} perpendicular to the disk. The misalignment in the case of 48~Per was first reported by \citep{Delaa2011a} and is discussed by \cite{Rivinius2013}, who caution that the small intrinsic polarization expected from 48~Per (due to its relatively small $i_\star$) makes it difficult to separate from the much larger interstellar contribution. Without uncertainties in the polarization position angles, it is hard to evaluate the statistical significance of these three (out of 11) apparently perpendicular angles. On one hand, disk asymmetries (caused by density waves) or contributions from other circumstellar emitting regions (for example, due to the presence of polar wind) could potentially change the simple prediction of the polarization position angle being strictly perpendicular to the disk. On the other hand, correction of observed polarization percentages and position angles for the interstellar contribution is highly non-trivial, as emphasized by \citet{Rivinius2013}. In the case of \citet{Yudin2001}, a map of the sky polarization in the observing field of each Be star, and its dependence on distance, was used to correct for the average effects of interstellar polarization; however, it is outside the scope of this paper to assess how well these corrections were performed and if further improvements in interstellar polarization corrections could be achieved.

Finally, we note that \cite{Cure2010} tested the distribution of polarization position angles from the catalogue of \citet{Yudin2001} to see if it was consistent with a uniform distribution, as one might expect for randomly oriented disks in the field Be star population. \cite{Cure2010} found that a K-S test rejected the null hypothesis of a uniform distribution, and suggested that the observed distribution is slightly bi-modal.

\section{Discussion}
\label{sec:discuss}

The viewing inclination of a Be star+disk system has long been known to play a major role in shaping the appearance of its H$\alpha$ emission line. However, the current work demonstrates {\it quantitatively\/} that fitting a {\it single\/} observed H$\alpha$ profile (${\rm SNR}=10^2$ and ${\cal R}=10^4$) to computed libraries allows an estimate of the system inclination to within $\pm10^\circ$, even though several disk density parameters must also be simultaneously determined. Direct comparison of these spectroscopically-determined inclinations to ones independently derived for sample of eleven Be stars using NPOI interferometric observations that spatially resolve their disks confirms the robustness and accuracy of the spectroscopic H$\alpha$ method; the differences in the viewing inclinations derived from these two methods are consistent with a Gaussian of zero mean and standard deviation $\sigma=7^\circ$.

Advantages of the H$\alpha$ spectroscopic method for determining viewing inclinations are manifest:\ only a single observed spectrum is required, as opposed to a time-series of observations for other methods. There is no obvious bias of the method to a particular range of inclinations, and the library-matching method naturally produces an uncertainty estimate for each derived inclination. While the method is applicable only to Be stars, these star+disk systems are sufficiently common that any open cluster young enough to contain main sequence B stars will also posses a significant population of Be stars. In addition, as Be stars are bright, they are detectable not only in Galactic open clusters, but also in the LMC/SMC and other members of the Local Group, allowing the effect of metallicity on open cluster spin-axes alignment to be explored. Finally, coupling the spectroscopic H$\alpha$ method to determine the viewing inclination with a continuum polarization measurement to constrain the in-sky component of the disk's (and hence star's) rotation axis, it should be possible to reconstruct the full 3-dimensional direction of the rotation axis of a Be star.

Going forward, we plan to further test our spectroscopic H$\alpha$ inclination angles by comparing to the large samples of Be star inclinations available from gravitational darkening modelling of Be star spectra \citep{Fremat2005a,Zorec2017}. We also plan to extend this method to modelling the H$\beta$ line in Be star spectra to test for consistency between the H$\alpha$ and H$\beta$ inclination results. Finally, we are attempting to extend our analysis to asymmetric H$\alpha$ profiles from Be stars that exhibit V/R variations by fitting the red and blue wings of the H$\alpha$ profile separately and looking for consistency between the inclination derived from the two profile halves.

\section*{Acknowledgements}
The authors would like to thank Anahi Granada and Dietrich Baade for helpful comments. T.\ A.\ A.\ S.\ acknowledges support from the Natural Sciences and Engineering Council of Canada through a Discovery Grant. C.\ T.\ acknowledges support from the National Science Foundation through grant AST-1614983. The Navy Precision Optical Interferometer is a joint project of the Naval Research Laboratory and the U.S. Naval Observatory, in cooperation with Lowell Observatory, and is funded by the Office of Naval Research and the Oceanographer of the Navy. We thank the Lowell Observatory for the telescope time used to obtain the H$\alpha$ spectra used in this work.


\appendix

\section{Limits on Interferometric Determinations of Inclination Angles}

To investigate the limitation of using the ratio of the minor-to-major axes as a proxy for inclination through $i_{\rm V^2}\equiv\cos^{-1}(b/a)$, we have generated synthetic, radiative transfer images for a typical Be star disk on the sky using the \texttt{Beray} code. These images were then used to generate interferometric visibilites that were analyzed using exactly the same pipeline and geometric models used in Section~\ref{sec:npoi}. The chosen model corresponded to a $5\,M_{\odot}$ model from Table~\ref{tab:centralB}, surrounded by $R_d=25\,R_{\odot}$ circumstellar disk with $\log_{10}\rho_0=-10.46$ and $n=2.00$. Synthetic images for inclinations $i_\star=10^\circ$ through $i_\star=90^\circ$ in steps of $10^\circ$ were computed in a 150~nm band-pass\footnote{The width of the band-pass was chosen to match the NPOI H$\alpha$-containing interferometric spectral channel.} centred on H$\alpha$. The images were Fourier transformed and then discretely sampled in the $(u,v)$ plane to form a set of synthetic ``observed" visibilities (see \citet{Sigut2015} for further details). A total of $N=200$ $(u,v)$ plane points were used, symmetrically placed between $k_{\rm min}=10 \times 10^6\;$cycles per radian and $k_{\rm max}=150 \times 10^6\;$cycles per radian. Visibility data were generated for 10 viewing inclinations, assuming a distance of $150\;$pc for the system, giving an angular diameter for the central star of $\theta_*\equiv 0.24\,$mas in Equation~\ref{eq:v2model}. A fractional uncertainty of 10\% was assumed for the simulated visibility data, and the uncertainties in the best-fit parameters $(a,b,c_*,\phi)$ were obtained via bootstrap Monte Carlo simulation.

\begin{figure}
\centering
\includegraphics[width=0.4\columnwidth]{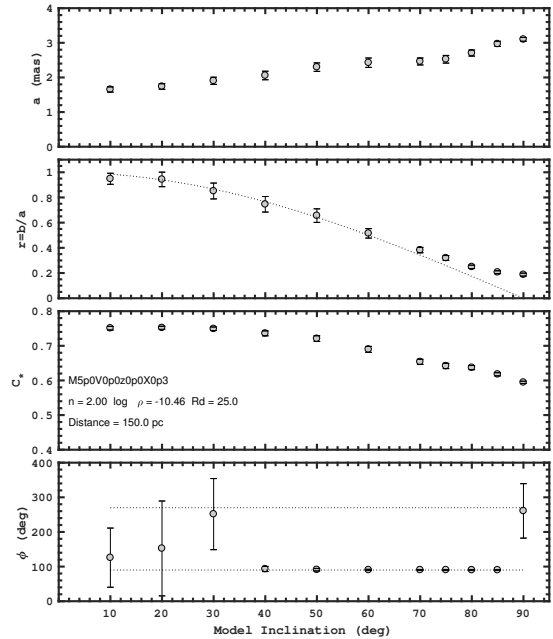}
\caption {Recovered parameters in the visibility model of Equation~\ref{eq:v2model} as a function of viewing inclination. The squared visibility fits are based on synthetic \texttt{Beray} images for a $5\,M_{\odot}$ star surrounded by $R_d=25\,R_{\odot}$ circumstellar disk with $\log_{10}\rho_0=-10.46$ and $n=2.00$. The panels are major axis $a$ (top panel, in mas), axial ratio $r=b/a$, stellar contribution to the light $c_*$, and disk position angle $\phi$ (bottom panel, in degrees). Each parameter is shown as a function of the inclination used to compute the image ($i$ in degrees). Error bars are 1$\sigma$ variation over 500 bootstrap Monte Carlo runs. In the axial ratio panel, the dotted line is the $r=b/a=\cos i$ prediction. In the bottom panel, the dotted lines are for $\phi=90^\circ$, the position angle of the computed image, and its compliment angle ($270^\circ$). \label{fig:v2fitparams}}  
\end{figure}

Results for the fitted model parameters $(a,b,c_*,\phi)$ are shown in Figure~\ref{fig:v2fitparams} as a function of the model inclination. Note that in all cases, the images generated by \texttt{Beray} used the same underlying stellar $(M,R,L)$ and disk $(n,\rho_0,R_d)$ model; however, the apparent major axis of the disk tends to increase for increasing inclination angle $i_\star$, even though $R_d$ is fixed. As $i_\star$ is increased, the oblique path through the disk allows the $\tau\approx 1$ to be achieved further from the star, resulting in a larger apparent disk. The recovered axial ratio $b/a$ is also shown and compared to the simple prediction $b/a=\cos i_\star$ where $i_\star$ is the inclination used to create the \texttt{Beray} image. Agreement between this simple prediction and the recovered parameter is very good and within the errors until $i_\star\approx 75^\circ$, when the data systematically lies above the prediction and outside the errors; this reflects the finite scale height of the disk. Finally, the recovered position angle is shown in the bottom panel; the images were computed for $\phi=90^\circ$ and either this value or its compliment ($\phi=270^\circ$) were recovered within the errors in all cases.  

\begin{figure}
\centering
\includegraphics[width=0.4\columnwidth]{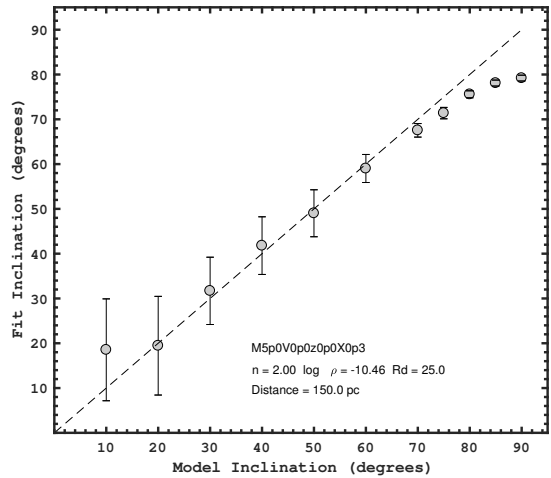}
\caption {Median inclination, $i_{\rm V^2}\equiv \cos^{-1}(b/a)$, plus 1$\sigma$ error, versus the actual inclination used in \texttt{Beray} to compute synthetic images for the same set of models as shown in Figure~\ref{fig:v2fitparams}. \label{fig:v2comp}}  
\end{figure}

Finally, Figure~\ref{fig:v2comp} shows the recovered inclination as $i_{V^2}\equiv \cos^{-1}(b/a)$ versus the model inclination actually used by \texttt{Beray} to compute the image. For smaller $i_\star$, $i_\star\le 20^\circ$ the uncertainties are large because the deviation from circular symmetry is small and strongly influenced by the assumed random errors. For $20^\circ < i_\star < 80^\circ$, the uncertainties are much smaller and very good estimates are recovered for the model inclinations. For $i_\star \ge 80^\circ$, the $\cos^{-1}(b/a)$ estimate underestimates the model inclination as the finite thickness of the disk provides a lower limit to the recovered minor axis $b$. However, even in these cases, it is clear from the fit that $i_\star\ge 80^\circ$.  

\vspace{0.1in}

\bibliography{citas.bib} 
\label{lastpage}
\end{document}